\documentclass[aps,pra,reprint,groupedaddress]{revtex4-1}
\usepackage[utf8]{inputenc}
\usepackage{ae,amsmath,amssymb,amsfonts,bbm,amsthm}
\usepackage[english]{babel}
\usepackage{braket}
\usepackage{graphicx}
\usepackage{calc}
\usepackage{color}
\usepackage[hidelinks]{hyperref}
\usepackage[caption=false]{subfig}
\usepackage[titletoc,title]{appendix}
\usepackage{placeins}
\usepackage{enumitem}
\usepackage{units}
\usepackage{here}

\newcommand{\ketb}[1]{\mathbf{\ket{#1}}}
\newcommand*\abs[1]{\left|#1\right|}
\newcommand\numberthis{\addtocounter{equation}{1}\tag{\theequation}}

\begin{document}

\title{Teleportation-Assisted Optical CSIGN Gates}

\author{Fabian Ewert}
\email[]{ewertf@uni-mainz.de}
\affiliation{Institute of Physics, Johannes Gutenberg-Universität Mainz, Staudingerweg 7, 55128 Mainz, Germany}
\author{Peter van Loock}
\email[]{loock@uni-mainz.de}
\affiliation{Institute of Physics, Johannes Gutenberg-Universität Mainz, Staudingerweg 7, 55128 Mainz, Germany}

\begin{abstract}
Reliable entangling gates for qubits encoded in single-photon states represent a major challenge on the road to scalable quantum computing architectures based on linear optics. In this work, we present two approaches to develop high-fidelity, near-deterministic controlled-sign-shift gates based on the techniques of quantum gate teleportation. On the one hand, teleportation in a discrete-variable setting, i.e., for qubits, offers unit-fidelity operations but suffers from low success probabilities. Here, we apply recent results on advanced linear optical Bell measurements to reach a near-deterministic regime. On the other hand, in the setting of continuous variables, associated with coherent states, squeezing, and, typically, Gaussian states, teleportation can be performed in a deterministic fashion, but the finite amount of squeezing implies an inevitable deformation of a teleported single-mode state. Using a new generalized form of the nonlinear-sign-shift gate for gate teleportation, we are able to achieve fidelities of the resulting \textsc{csign} gate above $90\%$. A special focus is also put on a comparison of the two approaches, not only with respect to fidelity and success probability, but also in terms of resource consumption.
\end{abstract}

\maketitle

\section{Introduction}
\label{sec:motivation}
In 2001, Knill, Laflamme and Milburn (KLM) showed that, quite surprisingly at that time, linear optical elements such as beam splitters and phase shifters are sufficient for universal quantum computation \cite{KLM}. They utilized measurements involving additional optical modes and photons to introduce an effective nonlinearity, i.e., the so-called nonlinear-sign-shift (\textsc{nss}) gate, to circumvent the need for a direct nonlinear interaction between photons. Using two \textsc{nss} gates together with beam splitters allows the construction of a linear optical entangling gate, the \textsc{csign} gate.

Unfortunately, this comes at the price that the \textsc{nss} gate, and therefore the \textsc{csign} gate as well, are probabilistic since the \textsc{nss} gate only succeeds if the measurements in the additional modes yield a certain result. More specifically, the \textsc{nss} gate succeeds, under perfect circumstances, only in one quarter of all cases. The resulting success probability of $1/16$ for the \textsc{csign} gate prevents its immediate use in large-scale quantum computation algorithms where hundreds or more of these gates are required for a reliable operation.

In order to circumvent this problem, KLM proposed to outsource the probabilistic application of the gates onto a suitable ancillary resource state and then perform quantum teleportation of the online qubits using that state. This procedure is commonly referred to as quantum gate teleportation \cite{Chuang}. Since the resource state contains none of the valuable quantum information processed in the computation algorithm, the probabilistic nature of the \textsc{nss} and \textsc{csign} gate can be counteracted by applying it to a sufficiently large number of resource states and only using those for the teleportation where the gate application succeeded. The success probability of the gate on the online-qubits is then simply given by the success probability of the teleportation. However, in the context of linear optical processing of qubits this teleportation success probability is determined by the efficiency of the required Bell measurement which in turn is usually limited to $1/2$ \cite{CalsamigliaNL}. In the last few years multiple methods to surpass this notorious $1/2$-limit have been proposed \cite{Grice,Zaidi,Ewert} and applying these to increase the gate teleportation success probability is very straightforward (see Sec.~\ref{sec:DVGT}).

In this work, however, we also investigate a different approach to a gate-teleportation-assisted \textsc{csign} gate. Instead of using the discrete-variable (DV) gate teleportation to teleport the entire \textsc{csign} gate, we propose to use a continuous-variable (CV) scheme to teleport the \textsc{nss} gate onto DV states (see Secs.~\ref{sec:CVQT} and \ref{sec:CVGT}). The main advantage there is that the CV teleportation \cite{Vaidman94,Braunstein98} is, in principle, deterministic and therefore removes the obstacle of low success probabilities for quantum computation. Moreover, the CV scheme relies on homodyne detections with near-unit efficiencies and, in particular, does not require photon-number resolving detectors. Yet, this comes at the price of a reduced fidelity of the gate, i.e., the teleported state is deformed, depending on the finite squeezing level of the suitably adapted two-mode squeezed resource state. We use conditioning of the homodyne detection within the CV teleportation to induce a trade-off between success probability and fidelity in the hope of finding an optimal point, where the fidelity is still very close to one and high success probabilities are achieved.
Nonetheless, we find that in order to increase both fidelity and success probability, further adjustments in the CV teleportation scheme are necessary. First, the standard correction operation (a displacement) is replaced by a nonlinear displacement, which can only be avoided entirely in the limit of conditioning onto zero-quadrature values for the homodyne-based CV Bell measurement (i.e., in the case of zero success probability). Second, the offline \textsc{nss} gate must be generalized in order to more closely resemble the action of a strong self-Kerr effect.

In Sec.~\ref{sec:comparison} we compare the two different approaches to teleportation-assisted \textsc{csign} gates (DV and CV) not only with respect to success probability and fidelity, but also in terms of resource consumption. While we find that in most situations the use of the advanced Bell measurement techniques in the DV scheme is more efficient, we still note that under certain circumstances the new CV scheme may be a valuable option.

\section{Linear optical CSIGN gates}\label{sec:linoptCSIGN}

Let us first shortly review the results of KLM~\cite{KLM} regarding the use of additional optical modes and measurements to create a linear-optics \textsc{csign} gate. Figure~\ref{fig:KLMNSS} shows a realization of the nonlinear-sign-shift gate equivalent to the one devised by KLM. Together with the input state $\ket{\text{in}}$ the ancillary states $\ket{1}$ (the one-photon Fock state) and $\ket{0}$ (the vacuum state) enter an array of beam splitters [with reflection amplitudes $r_1 = r_3=\cos(\pi/8)$ and $r_2=\tan(\pi/8)$] that realizes the following unitary transformation of the three mode creation operators:
\begin{align}
	\begin{pmatrix}
		a_1^\dagger\vphantom{\sqrt{\frac{3}{2}}}\\a_2^\dagger\\a_3^\dagger\vphantom{\sqrt{\frac{3}{2}}}
	\end{pmatrix}
	\rightarrow \begin{pmatrix}
		1-\sqrt{2} & \frac{-1}{\sqrt[4]{2}} & \sqrt{\frac{3}{\sqrt{2}}-2}\\
		\frac{-1}{\sqrt[4]{2}} & \frac12 & \frac{1}{\sqrt{2}}-\frac12\\
		\sqrt{\frac{3}{\sqrt{2}}-2} & \frac{1}{\sqrt{2}}-\frac12 & \sqrt{2}-\frac12
	\end{pmatrix}
	\begin{pmatrix}
		a_1^\dagger\vphantom{\sqrt{\frac{3}{2}}}\\a_2^\dagger\\a_3^\dagger\vphantom{\sqrt{\frac{3}{2}}}
	\end{pmatrix}.
\end{align}

Conditioned on the outcome of the photon detection in modes $2$ and $3$ being ``1" and ``0" respectively, a sign shift is induced on the two-photon Fock-state term of the input $\ket{\text{in}}$:
\begin{align}
	c_0\ket{0} + c_1\ket{1} + c_2\ket{2} \stackrel{\textsc{NSS}}{\longrightarrow} c_0\ket{0} + c_1\ket{1} - c_2\ket{2}. \label{eq:NSSaction}
\end{align}

Several important properties of this nonlinear-sign-shift gate should be noted. First, by construction this gate is probabilistic, since it will succeed only in the case that the correct detector clicks occur in the ancillary modes. A rather straightforward calculation shows that this happens only in one quarter of all cases, as long as the input state consists only of combinations of Fock states with less than three photons [as is assumed in Eq.~\eqref{eq:NSSaction}]. In the standard application of the \textsc{nss} gate, namely the linear-optics \textsc{csign} gate, which we will discuss below, the input to the \textsc{nss} gate is indeed limited to the lowest three Fock states, but, of course, it is also possible to apply it to higher Fock states. Especially in the context of the CV gate teleportation of a \textsc{nss} gate, which is discussed in Sec.~\ref{sec:CVGT}, this becomes important. As our second observation, we therefore find that the gate (conditioned in the same way as above) acts on an arbitrary Fock state as
\begin{align}
	\ket{n} \stackrel{\textsc{NSS}}{\longrightarrow} \frac{1}{2}(1-\sqrt{2})^n[1-(2+\sqrt{2})n] \ket{n},\label{eq:NSSactionN}
\end{align}
which shows that, while the \textsc{nss} gate is diagonal in the Fock basis, it is not unitary but exponentially suppresses higher Fock states. It should be noted that, different from Eq.~\eqref{eq:NSSaction}, the output state is not yet normalized as otherwise the exponential suppression would remain hidden.

\def\scalea{0.9}
\begin{figure}
\centering
\subfloat[\textsc{nss}\label{fig:KLMNSS}]{\includegraphics[scale=\scalea]{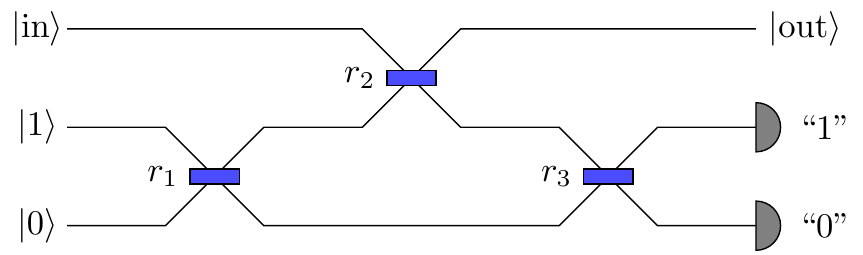}}\newline
\subfloat[\textsc{csign}\label{fig:KLMCZ}]{\includegraphics[scale=\scalea]{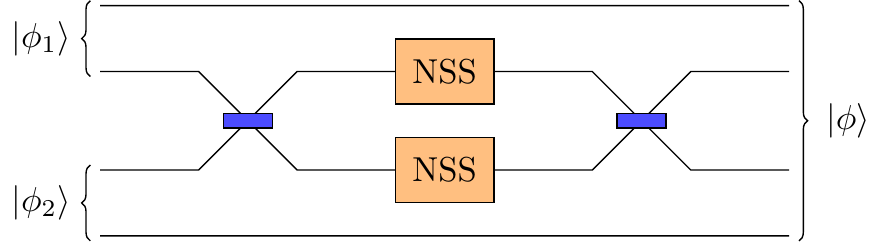}}
\caption{Nonlinear-sign-shift gate and linear-optics \textsc{csign} gate as proposed by KLM \protect\cite{KLM}.}\label{fig:KLMab}
\end{figure}

Let us now turn to the most important application of the \textsc{nss} gate, the construction of a linear-optics \textsc{csign} gate. A \textsc{csign} gate is defined on an arbitrary two-qubit state via
\begin{align}
	&\alpha \ketb{00} + \beta \ketb{01} + \gamma \ketb{10} + \delta \ketb{11}\nonumber\\
	\stackrel{\textsc{CSIGN}}{\longrightarrow} &\alpha \ketb{00} + \beta \ketb{01} + \gamma \ketb{10} - \delta \ketb{11}, \label{eq:CSIGNaction}
\end{align}
where we use $\ketb{0}=\ket{01}$ and $\ketb{1}=\ket{10}$ to denote the two dual-rail qubits. In dual-rail encoding, one of the most common encodings for qubits in optical quantum information processing, a \textsc{csign} gate can be realized with 50:50 beam splitters and two \textsc{nss} gates quite easily (see Fig.~\ref{fig:KLMCZ}). Since each of the dual-rail qubits consists only of one photon, the number of photons entering the \textsc{nss} gates in the setup of Fig.~\ref{fig:KLMCZ} is limited to two. Equation~\eqref{eq:NSSaction} then shows that the \textsc{nss} gate leaves the input unchanged unless exactly two photons enter a \textsc{nss} gate. This, however, can only occur if both photons are initially in the inner modes (note that for illustration the modes representing a logical one or zero in the dual-rail encoding are swapped in $\ket{\phi_1}$). In that case the Hong-Ou-Mandel effect \cite{HOM87} ensures that both photons are directed to the same \textsc{nss} gate and thus pick up a phase of $(-1)$, while the second beam splitter ensures that the output state is again in the two-qubit space with one photon for each qubit.

While this realization of a \textsc{csign} gate has been an enormous step in the direction of optical, universal quantum computation, as it is an entangling gate based purely on linear optics, its biggest drawback is that it inherits the probabilistic nature of the \textsc{nss} gate. For the \textsc{csign} gate to operate successfully, both \textsc{nss} gates have to succeed, which leads to a total success probability for the \textsc{csign} gate of $1/16$. In 2002 Knill \cite{Knill} proposed a \textsc{csign} gate based on the same idea of measuring additional modes and photons. This realization achieves a success probability of $2/27$ using less resources (fewer modes and detectors) than that based on the original \textsc{nss} gate. However, for large-scale quantum computing algorithms, where many \textsc{csign} gates may be required, this success probability is still small, thus posing a major problem. KLM therefore proposed to use a combination of quantum gate teleportation and quantum error correction codes in order to boost the success probability of the \textsc{csign} gate to a near-deterministic regime. Unfortunately, the resource requirements, especially those that stem from the need for a high-efficiency teleportation scheme, are rather big. In this work, our focus lies distinctively on using more advanced teleportation schemes in order to achieve high success probabilities for the \textsc{csign} gate.

\section{Discrete-variable gate teleportation}\label{sec:DVGT}

\begin{figure}[t]
	\centering
	\includegraphics[scale=1]{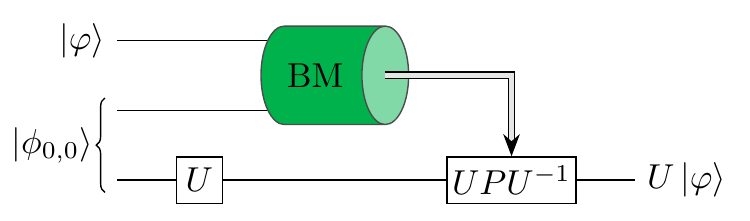}
	\caption{Gate teleportation of an unknown state $\ket{\varphi}$.}
	\label{fig:DVGateTele}
\end{figure}

The basic idea of quantum gate teleportation is depicted in Fig.~\ref{fig:DVGateTele}. Instead of applying a (probabilistic) gate directly to the online qubit, thus risking failure and thereby loss of the valuable quantum information, an ancillary Bell state, say $\ket{\phi_{0,0}} = \frac{1}{\sqrt{2}}(\ketb{00}+\ketb{11})$, is supplied (its generation may also be probabilistic) and the desired gate $U$ is performed on one of its qubits. Only in the case of heralded success of both state generation and gate application, the resulting state is used as resource for teleportation of the online qubit. As usual for teleportation schemes, the outcome of the Bell measurement determines which correction operation must be performed in order to obtain the desired output state, i.e., the online qubit with the gate $U$ applied to it. While for standard quantum state teleportation (which corresponds to $U=\mathbbm1$) the correction is a simple Pauli gate $P \in \{\mathbbm1, X, iY, Z\}$, in the case of gate teleportation the required correction is given by $U P U^{-1}$. Fortunately, there is an entire group of quantum gates $U$, called the Clifford group, for which $U P U^{-1}$ is itself a Pauli gate if $P$ is a Pauli gate, and the \textsc{csign} gate lies in this group (note that here and in Fig.~\ref{fig:DVGateTele} we focus on the teleportation of single-qubit gates, but the arguments can easily be extended to two-qubit gates by using two teleportation schemes).

The success probability $p_{\textrm{DV}}$ of applying the \textsc{csign} gate to the online qubit via gate teleportation is then given as the product of the probability of correctly creating the teleportation resource state
\begin{align}
	\ket{\textrm{res}} &= \textsc{csign}_{2,3} \ket{\phi_{0,0}}_{1,2} \ket{\phi_{0,0}}_{3,4}\nonumber\\
	&= \frac{1}{2}\left(\ketb{0000}+\ketb{0011}+\ketb{1100}-\ketb{1111}\right), \label{eq:DV_res_state}
\end{align}
which we denote $p_\textrm{sc,DV}$, and the probability of successfully performing two teleportations, which is determined by the Bell measurement success probability and thus given as $p_{BM}^2$. Both of these success probabilities depend upon the resources allocated to each task.

In order to compare several state creation approaches as well as various Bell measurement schemes, here we consider a single use of a single-photon source as the elementary unit of resource consumption. The cost of every operation in the gate teleportation schemes, e.g., application of the probabilistic \textsc{csign} gate, creation of Bell states or other states required for advanced Bell measurement schemes, will be reduced to this elementary unit. The same holds for operations in the CV setting, where the cost of all but one of the operations can be reduced to single-photon sources, and the contribution of the extra operation to the total cost is nearly negligible (see Sec.~\ref{sec:CVGT}). Therefore, single-photon sources provide a suitable figure of merit to compare the cost of the different approaches.

Let us now turn to the analysis of the success probabilities of state creation and Bell measurement and of their respective resource costs. We assume here that, since the gates are to be used in a (purely) optical quantum computing device, neither the online qubit nor the ancillary resource states are stored over prolonged periods of time. Thus, multiple attempts to creating the ancillary states are performed simultaneously to achieve a high gate teleportation success probability. Here we consider two different ways of creating the teleportation resource state $\ket{\textrm{res}}$ from single photons.

\begin{figure}
	\centering
	\includegraphics[width=0.8\columnwidth]{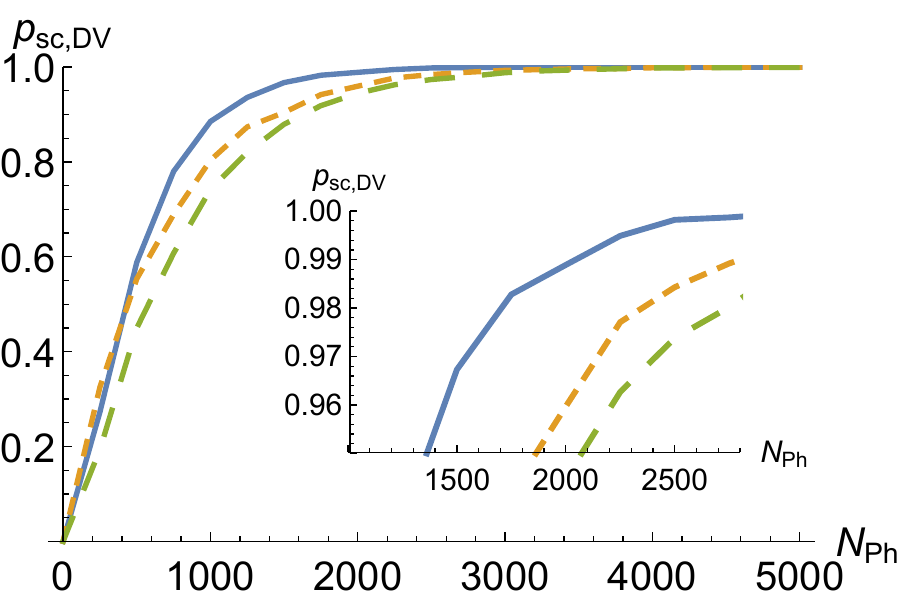}
	\caption{Success probability vs. number of required photons for the creation of the teleportation resource state $\ket{\textrm{res}}$ via Bell state creation followed by \textsc{csign} (orange, dashed), via connecting three-qubit clusters with the standard Bell measurement (green, widely dashed), or via connecting the cluster states with the advanced Bell measurement of Ref.~\protect\cite{Ewert} (blue, solid).}
	\label{fig:DV_res_creation}
\end{figure}

First, the most straightforward way is to generate (in a heralded fashion) two Bell states $\ket{\phi_{0,0}}$, which is possible with a success probability of $3/16$ from four single photons \cite{Zhang08}, followed by application of a probabilistic \textsc{csign} gate. The most resource efficient of the available linear optical \textsc{csign} gates is the one presented by Knill in Ref.~\cite{Knill} which utilizes two single photons to achieve a success probability of $2/27$. Assuming the use of multiplexing, i.e., successfully generated Bell states are directed to the \textsc{csign} gates such that no or at most one Bell state is wasted, we find that on average
\begin{align}
	n_\textrm{Knill} = \frac{27}{2} \cdot 2 + \frac{27}{2} \cdot 2 \cdot \frac{16}{3} \cdot 4 = 603 \label{eq:nKnill}
\end{align}
photons are required to successfully create one instance of $\ket{\textrm{res}}$. The actual relation between single-photon sources used and success probability obtained can easily be determined via a simple Monte-Carlo simulation. The resulting function is plotted in Fig.~\ref{fig:DV_res_creation}.

A second variant to creating the state $\ket{\textrm{res}}$ is based on the observation that an actual application of the \textrm{csign} gate to the resource state is not mandatory. Any state creation scheme that produces the state $\ket{\textrm{res}}$ is sufficient. Since $\ket{\textrm{res}}$ is a linear four-qubit cluster state \cite{RaussendorfBriegel}, it can be ``grown" from three-qubit cluster states via Bell measurement (see, for example, Appendix D of Ref.~\cite{EwertPRA}). The linear three-qubit cluster states, corresponding to three-qubit GHZ (Greenberger-Horne-Zeilinger) states, can be created from six single photons with a probability of $1/32$ \cite{Varnava08}, whereas the Bell measurement usually is limited to a success probability of $1/2$ in linear optics \cite{CalsamigliaNL}. A calculation analogous to that in Eq.~\eqref{eq:nKnill} gives an average consumption of $n_\textrm{cluster} = 768$ photons for one instance of $\ket{\textrm{res}}$. The results of the Monte-Carlo simulation (see Fig.~\ref{fig:DV_res_creation}) also show that this approach is less efficient than that based on Knill's \textsc{csign} gate. However, in recent years, several schemes have been proposed that achieve Bell measurement success probabilities significantly higher than $1/2$ with linear optical tools \cite{Grice, Zaidi, Ewert}. The most resource efficient one (again in terms of single-photon sources) in our setting of connecting two three-qubit clusters is that presented in Ref.~\cite{Ewert} where four additional photons allow for an increase of the Bell measurement success probability to $3/4$. With this advanced Bell measurement scheme the average number of photons required for one instance of $\ket{\textrm{res}}$ is reduced to $n_\textrm{cluster,adv} = 517$ which corresponds to a cost reduction of about $15\%$. Figure~\ref{fig:DV_res_creation} shows that when the goal is a high state generation rate, the saving of single-photon sources is even more substantial. To achieve a state creation probability of $0.95$, the advanced cluster-based scheme requires approximately $30\%$ less photons than that based on the application of Knill's \textsc{csign} gate.

As mentioned above, the creation of the state $\ket{\textrm{res}}$ represents only one part of the probabilistic nature of gate teleportation. We shall now focus on the success probability of the Bell measurements. In their seminal paper, KLM already proposed a method to increase the teleportation success probability by replacing the Bell measurement with a more sophisticated measurement based on an $(n+1)$-mode Fourier transformation. However, this also includes a more complicated resource state than $\ket{\textrm{res}}$, whose creation involves the application of multiple \textsc{csign} gates in such a fashion that even with multiplexing at several stages the average number of photons required to obtain even the simplest of the advanced resource states is in the range of millions, thus making this approach rather impractical. Nielsen~\cite{Nielsen} as well as Browne and Rudolph~\cite{BrowneRudolph} have already shown that based upon the concept of cluster state computation introduced by Raussendorf and Briegel~\cite{RaussendorfBriegel} much more resource-efficient quantum computation architectures with linear optics are possible. In this work, however, we stick to the idea of quantum gate teleportation and aim to increase the success probability of the gate teleportation to a near deterministic regime. To this end, again, the advanced Bell measurement schemes in Refs.~\cite{Grice,Zaidi,Ewert} are utilized.

While the Bell measurements of Refs.~\cite{Grice} and \cite{Ewert} use ancillary states for their advanced Bell measurements, in Ref.~\cite{Zaidi} single-mode squeezing is employed as a resource. The cost of this squeezing cannot be represented in terms of single-photon sources in a straightforward manner. Moreover, as the squeezing-based Bell measurement cannot exceed a success probability of $64.3\%$, it is not suitable for near-deterministic teleportation anyway. We thus concentrate on those schemes based on ancillary photons. Of these two, the scheme in Ref.~\cite{Ewert} achieving $p_\textrm{BM}=3/4$ with four single unentangled photons is most resource-efficient, but an explicit linear optical setup to create the ancillary states required in Ref.~\cite{Ewert} for $p_\textrm{BM}\longrightarrow 1$ from single photons is not known. On the other hand, the ancillary states in Ref.~\cite{Grice} are Bell and GHZ states of increasing size and as such cluster states that can be generated from single photons via first creating three-qubit linear cluster states and then joining these via Bell measurements. Therefore, this scheme will serve as our representative for the DV gate teleportation of \textsc{csign}. Nevertheless, in view of the many similarities between the schemes of Refs.~\cite{Grice} and \cite{Ewert}, it seems reasonable to assume that the ancillary states of Ref.~\cite{Ewert} may also be producible at a comparable, if not even smaller cost (like for the case $p_\textrm{BM}=3/4$).

The ancillary states of Ref.~\cite{Grice} that allow for a Bell measurement with $p_\textrm{BM} = 1-2^{-N}$ are of the form
\begin{align}
	\ket{\phi_{0,0}} \ket{\textrm{GHZ}_4} \ket{\textrm{GHZ}_8} \dots \ket{\textrm{GHZ}_{2^{N-1}}},\label{eq:Grice_anc}
\end{align}
where $\ket{\textrm{GHZ}_j} = \frac{1}{\sqrt{2}} \left( \ketb{0}^{\otimes j} + \ketb{1}^{\otimes j} \right)$, and one may note $\ket{\phi_{0,0}}=\ket{\textrm{GHZ}_2}$. As mentioned above, GHZ states can be combined using a Bell measurement \cite{BrowneRudolph, Varnava08}. After suitable Pauli corrections, determined by the outcome of the Bell measurement, one finds
\begin{align}
	\ket{GHZ_{j}} \otimes \ket{GHZ_{k}} \stackrel{\rm BM}{\longrightarrow} \ket{GHZ_{j+k-2}}.\label{eq:GHZGHZ}
\end{align}
Especially for large GHZ states this means that with every Bell measurement the size of the GHZ state is almost doubled. This allows for a rough estimate of the required number of single photons: to create the largest GHZ state required, $\ket{\textrm{GHZ}_{2^{N-1}}}$, $N-1$ (for $N>3$) successful Bell measurement levels, each nearly doubling the size of the GHZ state, are required when starting with $\ket{\textrm{GHZ}_3}$ states. Each of the latter consumes on average $32\cdot6$ photons in its creaton \cite{Varnava08}, corresponding to a three-photon state conditioned upon detecting three photons and a success probability of $1/32$. With the Bell measurement ($p_\textrm{BM}=3/4$) of Ref.~\cite{Ewert} to join the GHZ states, four additional photons per Bell measurement must be added. Thus, we estimate that approximately $200 \left(\frac{8}{3}\right)^{N-1}$ photons are needed to create $\ket{\textrm{GHZ}_{2^{N-1}}}$. In its creation process, the remaining smaller GHZ states of the ancillary resource in Eq.~\eqref{eq:Grice_anc} can be extracted as by-products.

\begin{figure}
	\centering
	\includegraphics[width=0.8\columnwidth]{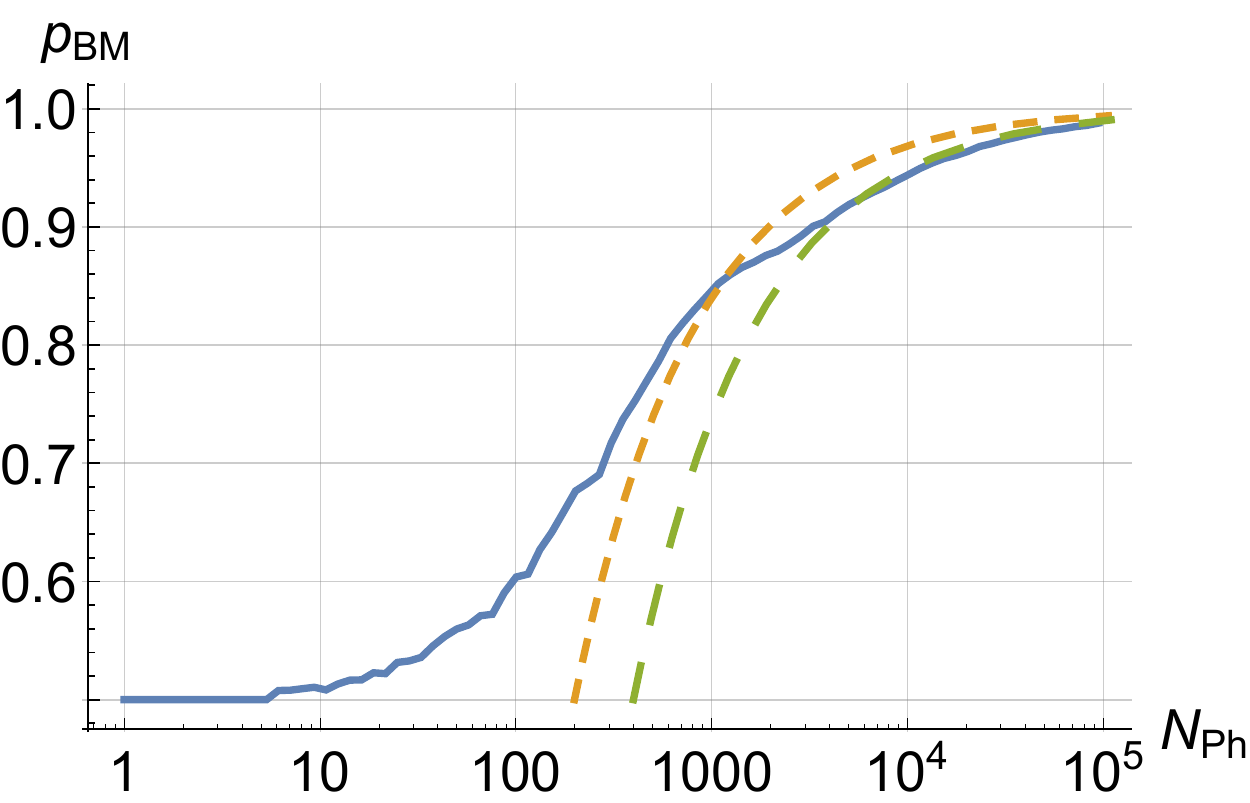}
	\caption{Success probability of the Bell measurement in Ref.~\protect\cite{Grice} vs. the number of required photons for the creation of the corresponding ancillary states. According to the Monte-Carlo simulation (blue, solid), the rough estimate presented in the main text (orange, dashed) gives a number for the required photon-sources that is too small by a factor of two (green, widely dashed) for large Bell measurement success probabilities, while it is far to high for $p_\textrm{BM}<0.7$.}
	\label{fig:DV_pBM}
\end{figure}

Figure~\ref{fig:DV_pBM} shows the results of a Monte-Carlo simulation that also includes the extraction of the additional GHZ states as well as the special behavior for $N\leq3$. We find that our estimate for the resource consumption is too small by almost exactly a factor of two for large numbers of single-photon sources (corresponding to values of $p_\textrm{BM}$ close to one), whereas for Bell measurement success probabilities only slightly above $1/2$ it is, as expected, far too large. It is worth noting that in the Monte-Carlo simulation we made use of a particular property of the Bell measurement scheme in Ref.~\cite{Grice}. If the creation of the largest GHZ state in Eq.~\eqref{eq:Grice_anc} fails, the remaining states can still be used for a Bell measurement which then succeeds with probability $1-2^{-(N-1)}$. The optical setup need not even be changed at all, as long as the ``failed" GHZ state is replaced by a vacuum input (see the appendix of Ref.~\cite{Grice}).

To summarize this section, both the creation of the teleportation resources state $\ket{\textrm{res}}$ and the Bell measurement required for teleportation can be performed near-deterministically with an overhead of a few thousand single-photon sources, which beats the original approach of KLM by several orders of magnitude. In Sec.~\ref{sec:comparison} these results are compared to the case of the CV-based gate teleportation presented in the following sections.

\section{Continuous-variable state teleportation}\label{sec:CVQT}

\begin{figure}
	\centering
	\includegraphics[width=0.7\columnwidth]{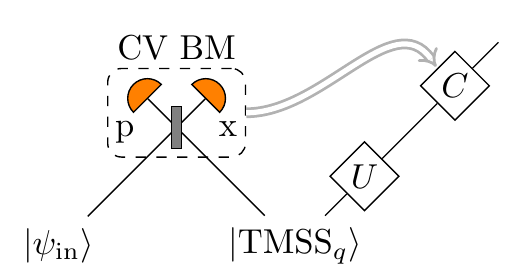}
	\caption{Continuous-variable teleportation. The input state $\ket{\psi_\textrm{in}}$ is combined with one half of a two-mode squeezed state (to which, in the case of gate teleportation, the gate $U$ is applied) at a beam splitter followed by homodyne detection. This is referred to as continuous-variable Bell measurement (highlighted area). Depending on the outcome $\beta$ of the CV Bell measurement (see the main text) a correction operation $C$ is applied. In the case of state teleportation, $C$ is a displacement $D(g\beta)$, where $g$ is the so-called gain-tuning parameter. In the case of gate teleportation, $C$ is usually chosen as $U D(g\beta) U^{-1}$.}
	\label{fig:CVGateTele}
\end{figure}

Quantum teleportation with continuous variables is, in principle, very similar to the DV case (see Fig.~\ref{fig:CVGateTele} and compare to Fig.~\ref{fig:DVGateTele}). On the state to be teleported and on one half of an entangled resource state a joint measurement is performed, projecting the other half of the resource onto the original state (after a suitable correction operation determined by the outcome of the joint measurement). However, since in the CV case the ancillary resource state is a two-mode squeezed state (instead of a Bell state for DV), and the joint measurement is performed via homodyne detections (replacing the DV Bell measurement), there are two major differences from the DV case. On the one hand, a CV Bell measurement based on homodyne detectors and beam splitters is, in principle, deterministic. This is one of the main motivations to investigate the possibility of an efficient quantum gate teleportation with continuous variables, as the small teleportation success probability is a big obstacle in the DV case when it is based on a standard linear-optics Bell measurement. Moreover, homodyne detectors still tend to have a higher quantum efficiency than photon detectors measuring Fock states. On the other hand, the two-mode squeezed state, while it is easy to create from relatively cheap resources and, most important here, deterministic in its production, can never become a maximally entangled-state resource like a DV Bell state, because the squeezing is always finite. Therefore, a teleportation with continuous variables will always, at least to some extent, deform the initial state in an undesired way. Throughout this work, we quantify this deformation via the fidelity between the actual output of the (gate) teleportation protocol and the desired output state, i.e., the input state (with the quantum gate applied to it directly).

Before we focus on the quantum gate teleportation, let us first introduce the formalism that we use here to generally describe CV teleportation. As both the input state of the teleportation and the gate to be teleported (the \textsc{nss} gate) are DV gates, we formulate our calculations in terms of Fock states. The two-mode squeezed state is given as
\begin{align}
	\ket{\textrm{TMSS}_q}=\sqrt{1-q^2}\sum_{n=0}^\infty q^n \ket{n,n},
\end{align}
where the strength of the squeezing, represented by $q$, is given as $-10 \log_{10}\left[e^{-2\operatorname{artanh}(q)}\right]$~dB. This state can rather easily be generated from two equally squeezed single-mode vacuum states (one position-squeezed and one momentum-squeezed) by combining them at a 50:50 beam splitter. The joint measurement depicted in Fig.~\ref{fig:CVGateTele}, which we refer to as CV Bell measurement, consists of a 50:50 beam splitter and homodyne detection of the (dimensionless) quadratures $\hat{x} = \frac{1}{2} \left(\hat{a}+\hat{a}^\dagger\right)$ in one mode and $\hat{p} = \frac{1}{2i} \left(\hat{a}-\hat{a}^\dagger\right)$ in the other one. This corresponds to a simultaneous measurement of the quadratures $\hat{x}_- = \hat{x}_1 - \hat{x}_2$ and $\hat{p}_+ = \hat{p}_1 + \hat{p}_2$ of the incoming (two-mode) state (prior to the beam splitter). It should be noted that, due to the beam splitter operation, the measurement results in the homodyne detections are actually $x =\frac{1}{\sqrt{2}} (x_1 - x_2)$ and $p = \frac{1}{\sqrt{2}}(p_1 + p_2)$. Thus, in order to obtain the desired values of $x_-$ and $p_+$ a multiplication with $\sqrt{2}$ is required. The final measurement result (after multiplication with $\sqrt{2}$) can conveniently be expressed as $\beta = x_-+ i p_+$ such that the whole measurement corresponds to a projection onto the state \cite{Hofmann01}
\begin{align}
	\ket{\beta_\textrm{hm}} = \frac{1}{\sqrt{\pi}} D_1(\beta) \sum_{n=0}^\infty \ket{n,n}.
\end{align}
Here $D_1(\beta)$ is the displacement operator on mode one and the factor $1/\sqrt{\pi}$ ensures that the probability of obtaining any measurement result is normalized to one. The state $\ket{\beta_\textrm{hm}}$ itself is not normalizable and unphysical. In reality, the resolution of the homodyne detectors is finite and thus the measurement operators of the CV Bell measurement do not exhibit such an unphysicality. However, the idealized state $\ket{\beta_\textrm{hm}}$ is a useful tool to describe both unconditional CV quantum teleportation and CV quantum teleportation conditioned upon a certain subset of the CV Bell measurement results (see Appendix~\ref{sec:AppA}). Taking into account the correction operation, i.e., the displacement $D(g \beta)$, where the so-called gain-tuning parameter $g$ represents a degree of freedom in the choice of the correction operation (it is typically chosen either $g=q$ or $g=1$), the action of the quantum state teleportation ($U=\mathbbm{1}$ in Fig.~\ref{fig:CVGateTele}) is given by the transfer operator
\begin{align}
	T_g(\beta) = \sqrt{\frac{1-q^2}{\pi}} D(g \beta) \sum_{n=0}^\infty q^n \ket{n}\!\bra{n} D^\dagger (\beta). \label{eq:transfer}
\end{align}
Note that the involved optical modes are not especially clarified for brevity. For further details, see, for example, Ref.~\cite{Hofmann01}.

\begin{figure}
	\centering
	\includegraphics[height=0.52\columnwidth]{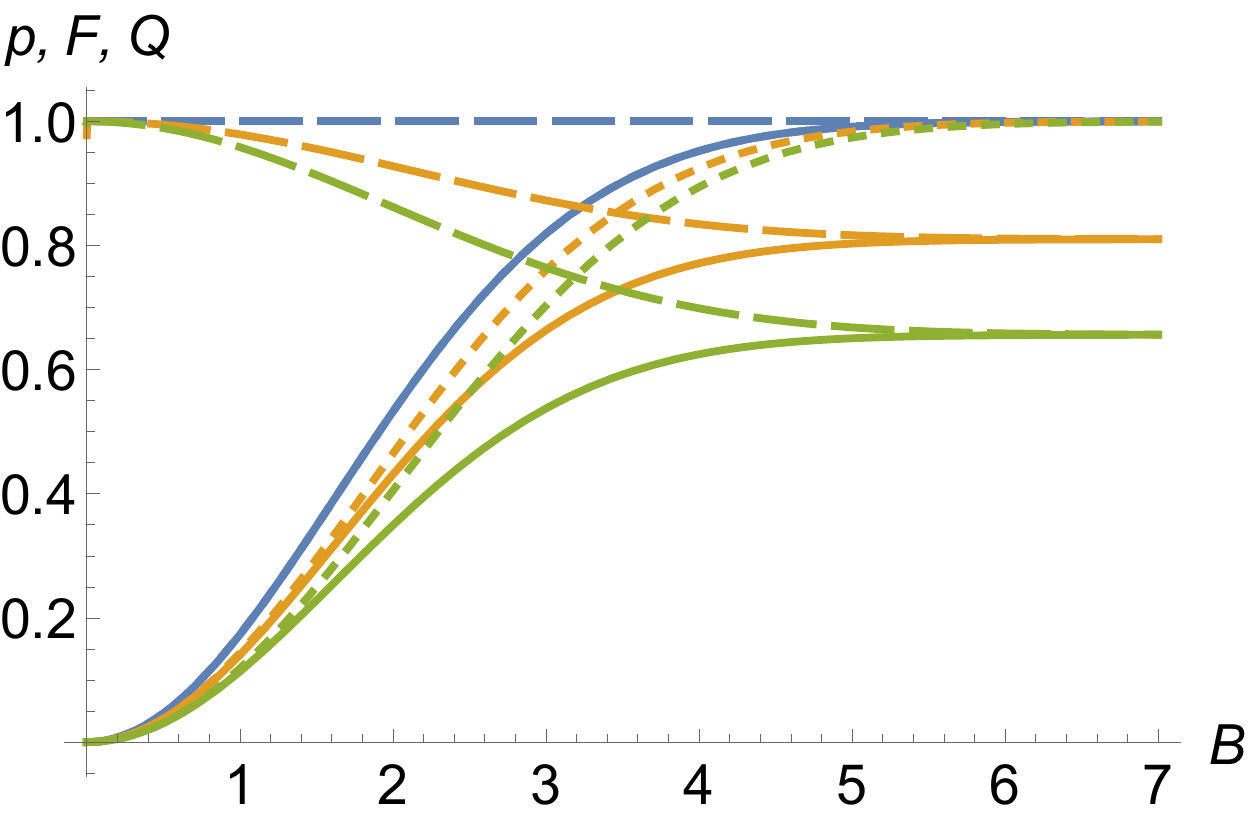}
	\caption{Trade-off between success probability (dotted) and fidelity (dashed) when conditioning the teleportation on a CV Bell measurement result of $\abs{\beta} \leq B$. Here we assume a squeezing of $q=0.9$, corresponding to $12.8$~dB, and we set $g=q$. The solid curves represent the quality $Q$ of the teleportation, as defined in the main text. From top to bottom (blue, yellow, green) the results for the Fock states $\ket{0}$, $\ket{1}$, and $\ket{2}$ are shown. For superpositions of these states, the corresponding curves always lie between the two extrema given by $\ket{0}$ and $\ket{2}$. In particular, this implies that the fidelity obtained for $\ket{2}$ represents the worst-case fidelity of the teleportation, when restricting the input to the first three Fock states. In any case, the quality increases monotonically with increasing conditioning radius $B$. Thus, not conditioning at all yields the highest quality. (The value of $q$ is chosen for the sake of visual distinguishability. With a squeezing of $q=0.96$, corresponding to $\sim 17$~dB which lies in the regime of the current experimental record \protect\cite{Schnabel16}, fidelities above $90\%$ are achieved.)}
	\label{fig:CV_tradeoff}
\end{figure}

With the transfer operator $T_g(\beta)$ it is possible to calculate both the probability (density) of obtaining the result $\beta$ and the fidelity for a given measurement result (the latter as a function of $g$) for a given input state (see Appendix~\ref{sec:AppA}). For large squeezing, $q\rightarrow 1$, we can infer from Eq.~\eqref{eq:transfer} that $T_{g=q}(\beta)\sim \mathbbm{1}$ corresponding to near-perfect teleportation. For a given finite squeezing value $q$, however, one quickly finds that the fidelity decreases quite substantially for increasing values of $\abs{\beta}$, but on the other hand, the probability of getting these large $\abs{\beta}$-values in the CV Bell measurement is rather small. It nonetheless seems sensible to condition the teleportation to smaller values of $|\beta|$ in order to achieve a higher fidelity on average. Of course, this comes at the price of sacrificing the deterministic teleportation for a probabilistic one. In Fig.~\ref{fig:CV_tradeoff} the behavior of both the success probability and the fidelity for the case of a quantum state teleportation ($U=\mathbbm{1}$) as a function of a conditioning parameter $B$ (the maximal value of $|\beta|$ to be accepted in the CV Bell measurement) is shown for the three Fock states $\ket{0}$, $\ket{1}$, and $\ket{2}$. For superpositions of these states, intermediate graphs are obtained, implying, in particular, that the fidelity obtained for $\ket{2}$ represents the worst-case fidelity of the teleportation, when restricting the input to the lowest three Fock states. We find that the product of success probability and fidelity, which we refer to as the ``quality" $Q=p\cdot F$ of the teleportation scheme, becomes maximal in the limit $B \rightarrow \infty$. This corresponds to not conditioning on the CV Bell measurement result at all. Of course, this observation does not exclude the possibility that conditioning may be advantageous in the gate teleportation setting. Nonetheless, while we have performed all calculations throughout this work with the conditioning parameter $B$ included, we found no instance where a conditioning with $B<\infty$ showed an improvement of the quality $Q$. Therefore, in the remainder, we shall focus on the unconditioned results. It should be kept in mind though that the quality, as defined above, is not the only possible figure of merit. For some applications a high fidelity might be extremely important and sacrificing (near-)unit success probability acceptable. In such a case, conditioning on the CV Bell measurement results provides a practical (and easily implementable) option.

\begin{figure}
	\centering
	\includegraphics[height=0.52\columnwidth]{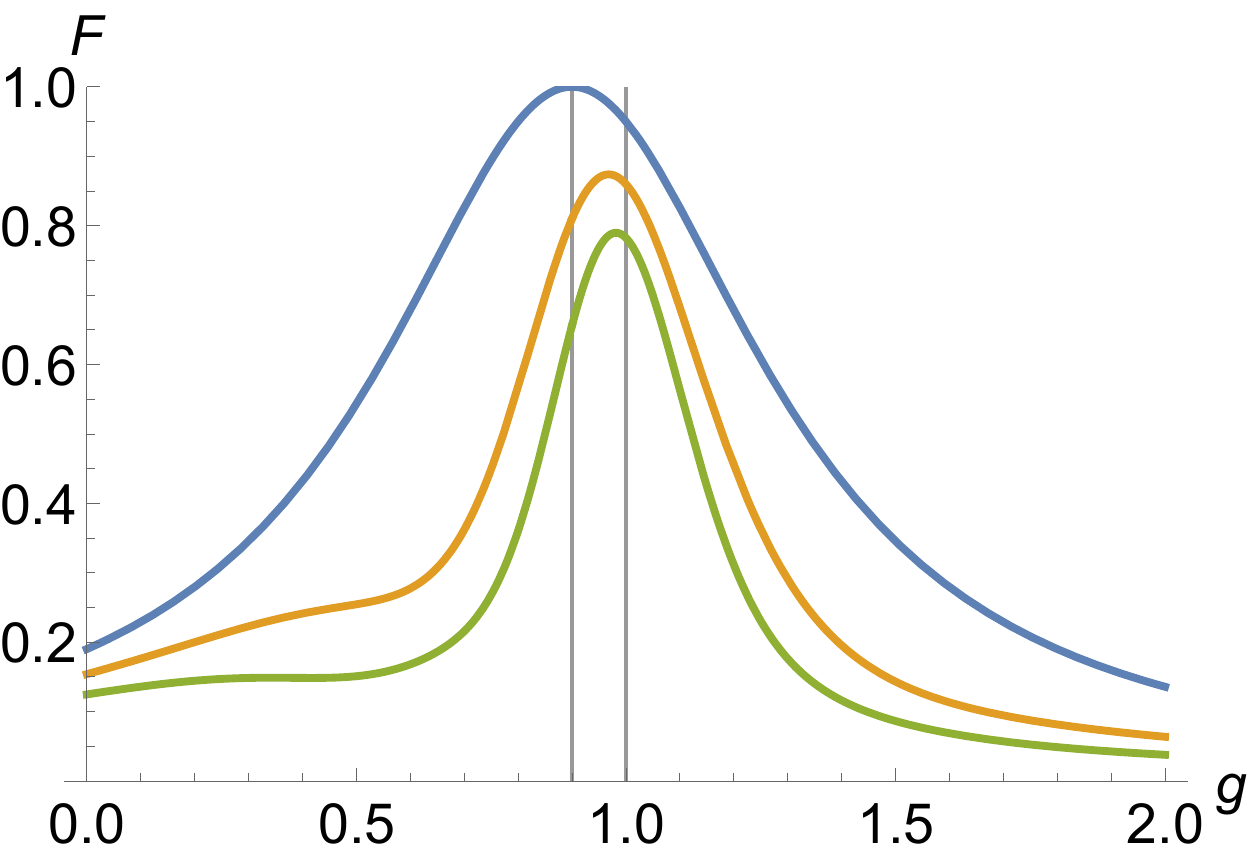}
	\caption{Fidelity of the (unconditioned) quantum state teleportation of the lowest three Fock states [from top to bottom $\ket{0}$ (blue), $\ket{1}$ (yellow), and $\ket{2}$ (green)] as a function of the gain-tuning parameter $g$ (with a squeezing of $q=0.9$, as in Fig.~\protect\ref{fig:CV_tradeoff}, chosen for visual distinguishability). Only for the vacuum state the maximal fidelity is obtained for $g=q$. The vertical lines are placed at $g=q$ and $g=1$.}
	\label{fig:CV_gain}
\end{figure}

In the above considerations, we have neglected the gain tuning parameter $g$, i.e., we set it to $g=q$, which is the standard choice besides unit gain. That choice is optimal for teleporting coherent states $\ket{\alpha}$, as with $g=q$ the output state in this case is the coherent state $\ket{q \alpha}$, independent of the measurement result $\beta$ (after normalization). More generally, for arbitrary input states, the choice $g=q$ ensures that the teleportation process is equivalent to a pure loss channel with transmission parameter $q$ \cite{Hofmann01}. For the CV teleportation of a single-mode DV state, however, it is not immediately clear that $g=q$ represents the optimal choice with respect to fidelity. Figure~\ref{fig:CV_gain} shows the fidelity of the (unconditioned) teleportation of the Fock states $\ket{0}$, $\ket{1}$, and $\ket{2}$ as a function of the gain tuning parameter $g$. As expected, $g=q$ represents the optimal choice for $\ket{0}$, but for the other Fock states, a value of $g$ closer to one is better suited. Since, in most situations, the input state is arbitrary and unknown, a general optimal choice of $g$ cannot easily be found. However, in some cases, the possible input states are limited. For example, when teleporting an arbitrary dual-rail qubit $\ket{\psi_\textrm{DR}} = c_0 \ket{01} + c_1 \ket{10}$ with two parallel CV teleportation schemes \cite{Takeda2013, furusawaBook}, we found that the overall fidelity as a function of the gain parameter is maximal for $g \approx 0.6 + 0.4 q$, independent of $c_0$ and $c_1$. (This result is obtained when the error spaces, especially the loss space, are kept in the output state, whereas a postselection onto the original dual-rail qubit code space, of course, yields unit fidelity in the case of $g=q$ corresponding to a  total output state $q \ket{\psi_\textrm{DR}}\bra{\psi_\textrm{DR}} + (1-q) \ket{00}\bra{00}$.) Presumably, this is due to the fact that from the point of view of one teleportation scheme its input is either $\ket{0}$ or $\ket{1}$ and not a superposition. Thus, the total fidelity would be the product of the fidelities of teleporting $\ket{0}$ and $\ket{1}$ separately. And indeed, this yields a very good approximation to the actual fidelity of the total process but one should remember that, in general, fidelities, unlike success probabilities, cannot simply be multiplied. Since the application of the \textsc{nss} gate teleportation within the \textsc{csign} gate of Fig.~\ref{fig:KLMCZ} resembles a similar situation of limited input types, we keep the general gain-tuning parameter $g$ and do not restrict ourselves to a specific choice.

\section{Continuous-variable gate teleportation}\label{sec:CVGT}

Let us now turn to the CV gate teleportation \cite{Bartlett03} of the \textsc{nss} gate. As usual, the gate to be teleported (\textsc{nss}) is applied to one half of the resource state ($\ket{\textrm{TMSS}_q}$) and, conditioned on the outcome of the joint measurement (the CV Bell measurement with result $\beta$), an appropriate correction is performed. Usually the correction is defined via $U D(g \beta) U^{-1}$, but in the case of \textsc{nss} (see Eq.~\eqref{eq:NSSactionN}) this leads to a transformation of the form
\begin{align}
	\sum_{n,m=0}^\infty \braket{n|D(g\beta)|m} (1-\sqrt{2})^{n-m} \frac{1-(2+\sqrt{2})n}{1-(2+\sqrt{2})m}  \ket{n}\!\bra{m}, \label{eq:nsscorr}
\end{align}
which does not appear to be any simpler with respect to realization than the \textsc{nss} gate itself. Therefore, it is not suitable as a correction operation in the teleportation scheme. We are thus looking for a correction operation that can be implemented reliably and applies the same correction as that in Eq.~\eqref{eq:nsscorr}, at least on the lowest three Fock states. The most promising candidate turns out to be the correction operation corresponding to teleportation of a strong self-Kerr gate $U_\textrm{SK} = \exp[i \frac{\pi}{2}\hat{n}(\hat{n}-1)]$, as the self-Kerr gate coincides with the \textsc{nss} gate when applied only to superpositions of the three lowest Fock states. (If the self-Kerr gate itself were available, we could, of course, use it directly in the \textsc{csign} gate of Fig.~\ref{fig:KLMCZ} instead of the \textsc{nss} gates. A possible realization of such a strong self-Kerr effect could be a decomposition into sufficiently many weak elementary CV gates \cite{LLoyd99,Sefi11,Sefi13,Miyata16,Marek17}.) The correction operation is then given as
\begin{align}
	C(g \beta) = U_\textrm{SK} D(g \beta) U_\textrm{SK}^\dagger,
\end{align}
and with the help of the operator identity $\operatorname{Ad}_{e^X} = e^{\operatorname{ad}_X}$, where $\operatorname{Ad}_{A}Y = A Y A^{-1}$ is the conjugation with $A$ and $\operatorname{ad}_{A}X = [A,X]$ is the adjoint endomorphism associated with the commutator, it can be rewritten as
\begin{align}
	C(g \beta) &= \exp\left[g \beta \hat{a}^\dagger (-1)^{\hat{n}}-g\beta^* (-1)^{\hat{n}} \hat a\right]\nonumber\\
	&= \exp\left[(-1)^{\hat{n}+1}g(\beta \hat{a}^\dagger +\beta^* \hat a)\right].\label{eq:fdefdisp}
\end{align}
Here we used $e^{\operatorname{ad}_{A}}Y = \sum_{k=0}^\infty \frac{1}{k!}\left[A,Y\right]_k$, with the concatenated commutator defined by $\left[A,Y\right]_k = \left[A,\left[A,Y\right]_{k-1}\right]$ and $[A,Y]_0=Y$, as well as $\left[i\frac{\pi}{2}\hat{n}(\hat{n}-1),\hat{a}\right] = (-i \pi \hat{n}) \hat{a}$. Equation~\eqref{eq:fdefdisp} represents a so-called $f$-deformed displacement operator, where in our case the function $f$ is given as $f(\hat{n}) = \exp(i \pi \hat{n})$ (for an introduction to $f$-deformed oscillators, coherent states and displacement operators see, for example, Ref.~\cite{Aniello2000}). Even though, to the authors' best knowledge, no realization of such an operation with optical means has been presented so far, the theory of $f$-deformed oscillators has been investigated quite thoroughly (see, for example, Refs.~\cite{Aniello2000,Filho96,Man'ko97,Perez16}). It thus seems much more reasonable to assume that an operation like $C(g \beta)$ will be available at some point, compared to that defined by Eq.~\eqref{eq:nsscorr}. In the following analysis, we will therefore assume that $C(g\beta)$ is available and can be performed in a deterministic fashion. When comparing the DV and CV schemes, we have to bear in mind this extra complication of the CV scheme.

Under the above assumptions, i.e., the \textsc{nss} gate is applied to the two-mode squeezed vacuum state successfully and the correction $C(g \beta)$ is performed perfectly, we find that the fidelity of the \textsc{nss} gate teleportation is rather low (approximately $25\%$ for teleportation of a $\ket{2}$ state) even after optimizing the choice of $q$ and $g$. Such a small fidelity is, of course, as much a hindrance for the implementation in large-scale quantum computing algorithms as the small probability of the \textsc{csign} gate by Knill \cite{Knill} in the DV case. We conjecture that the reason for this small fidelity is the misalignment on higher Fock states between the \textsc{nss} gate and the correction corresponding to the strong self-Kerr gate, since the application of the \textsc{nss} gate to the two-mode squeezed state means that it acts on higher Fock states as well. Thus, in order to increase the fidelity while keeping $C(g\beta)$ as the correction operation, we develop a generalization of the \textsc{nss} gate to resemble the action of the strong self-Kerr gate not only on the Fock states $\ket{0}$, $\ket{1}$, and $\ket{2}$, but on all Fock states up to $\ket{d}$. By generalization we mean that the gates described are realized by a linear optical setup, into which a finite number of single photons as ancillae are injected together with the input state, and whose success is heralded by a certain click pattern of photon-number-resolving detectors. The details of this gate are presented in Appendix~\ref{sec:AppB}. There we find that such a gate can be realized with $d-1$ ancillary photons and it will act on an arbitrary Fock state as
\begin{align}
	\ket{n} &\stackrel{\textsc{nss}_d}{\longrightarrow} \sqrt{p_d} \sum_{k=0}^{d-1} \sum_{l=0}^k \binom{n}{k} \binom{k}{l}\nonumber\\
	 & \qquad\left[(-1)^{d+1} \tan\left(\frac{\pi}{4d}\right)\right]^{n-k+l} U_\textrm{SK}(k+l) \ket{n},\label{eq:nssd}
\end{align}
where $U_\textrm{SK}(m) = \exp[i \frac{\pi}{2}m(m-1)]$, and $p_d$ is the success probability of the gate which depends on the actual optical setup used. Although not immediately perceivable from Eq.~\eqref{eq:nssd}, this gate resembles the action of the strong self-Kerr effect on all Fock states up to $\ket{d}$ (and superpositions of these), while the higher Fock states are suppressed exponentially. In our analysis we found setups where $p_d\approx 25\cdot 10^{-d}$ for $d\geq2$. As one may have suspected in advance, the success probability of these gates decreases exponentially with $d$. However, in the construction of the optical setups we have restricted the available degrees of freedom rather strictly in order to minimize the numerical expenses and to find a working proof of principle. Furthermore, in Ref.~\cite{Scheel05}, a very similar family of gates, i.e., another generalization of the \textsc{nss} gate, was developed. There, the phase shift occurs only on the highest Fock state, i.e.,
\begin{align}
	\sum_{k=0}^d c_k \ket{k} \rightarrow \sum_{k=0}^{d-1} c_k \ket{k} - c_d \ket{d},
\end{align}
and it was shown that this gate can be implemented with linear optics and a success probability of $1/d^2$. Due to the striking similarity to our own setting, it seems reasonable to assume that a realization for our generalization of the \textsc{nss} gate exists with the same scaling in the success probability. Therefore, in the discussion of the resource consumption, we will use this success probability as the most optimistic estimate.

\begin{figure}
	\centering
	\includegraphics[height=0.52\columnwidth]{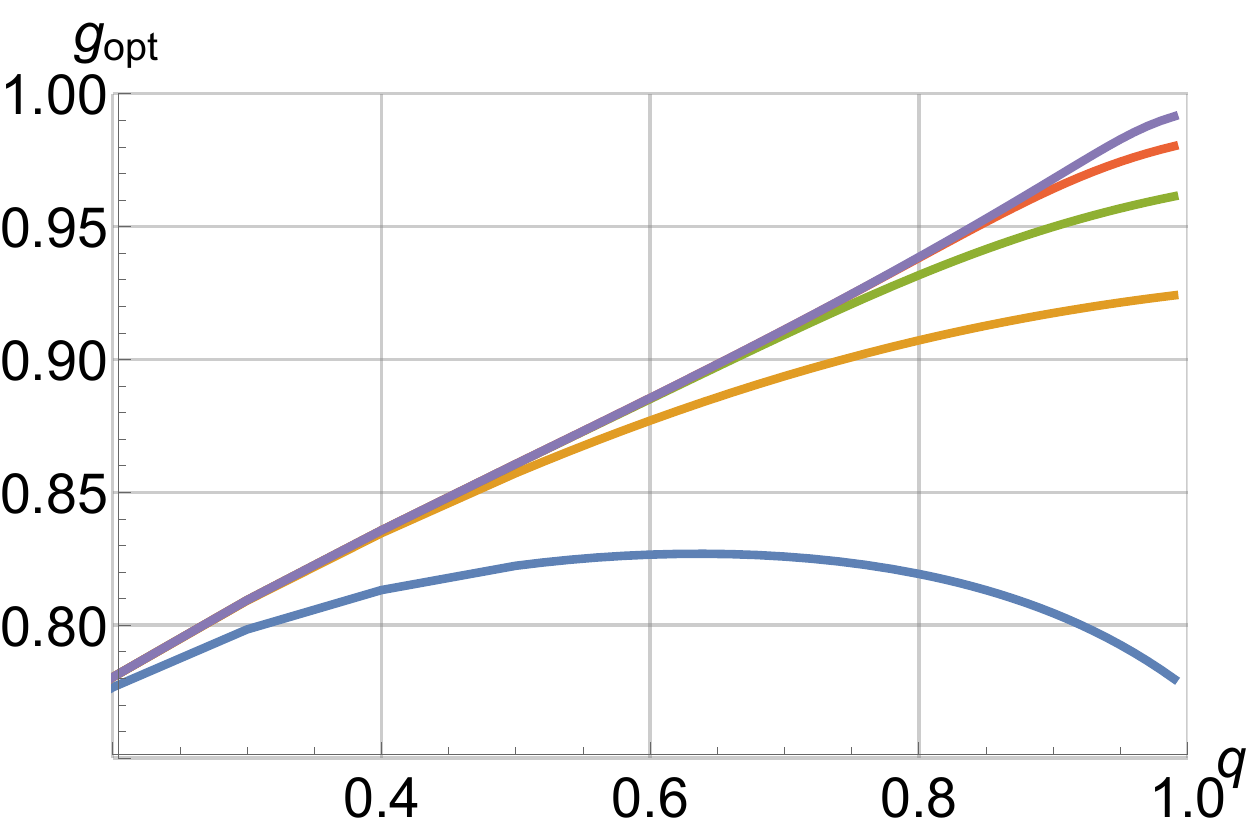}
	\caption{Optimal choice of the gain-tuning parameter $g$ for the two $\textsc{nss}_d$ gate teleportations in the \textsc{csign} gate as a function of the squeezing $q$. From bottom to top: $d=2,5,10,20,50$.}
	\label{fig:CZ_gopt}
\end{figure}

With the above generalization of the \textsc{nss} gate to $\textsc{nss}_d$, in combination with the $f$-deformed displacement of Eq.~\eqref{eq:fdefdisp} as correction operation, it is possible to achieve fidelities of the \textsc{nss} gate teleportation, and subsequently of the resulting \textsc{csign} gate, approaching unity, as long as $d$ is chosen sufficiently large. The derivation of explicit formulae for the fidelities is outlined in Appendix~\ref{sec:AppC}. The main idea to obtain numerically tractable equations is that, since the $\textsc{nss}_d$ gates suppress Fock states $\ket{n}$ with $n>d$ exponentially, it suffices to model the new teleportation resource state as
\begin{align}
	\mathcal{N}_{d,t}\ (\mathbbm{1}\otimes\textsc{nss}_d) \sum_{n=0}^{d+t} q^n \ket{n,n}, \label{eq:cvres}
\end{align}
i.e., the two-mode squeezed state is truncated ($\mathcal{N}_{d,t}$ is the new normalization constant). In fact, in all our calculations we found that for $t\geq2$ the fidelity becomes basically independent of $t$, implying that only the two Fock states directly above $\ket{d}$ have a substantial (negative) impact. Furthermore, as was already observable in the case of CV state teleportation (Sec.~\ref{sec:CVQT}), the lowest fidelities for teleportation are obtained when the input state is the Fock state $\ket{2}$. Thus, we easily obtain the worst-case fidelity of the \textsc{nss} gate teleportation. For the case of the total \textsc{csign} gate (constructed according to Fig.~\ref{fig:KLMCZ} with two gate-teleported \textsc{nss} gates), the worst-case fidelity is hence assumed when applying the gate to the state $\ketb{11}$ [see Eq.~\eqref{eq:CSIGNaction}]. This facilitates the computation of the fidelity and also allows for an optimization of the gain-tuning parameter $g$, analogous to the case of the dual-rail-qubit teleportation described in Sec.~\ref{sec:CVQT}. The optimal choice of $g$ as a function of $q$ in order to maximize the worst-case fidelity of the \textsc{csign} gate is depicted in Fig.~\ref{fig:CZ_gopt} for various $\textsc{nss}_d$ gates ranging from $d=2$ to $d=50$. For large values of $d$, i.e., for very close approximations to the self-Kerr gate, we obtain, as in the dual-rail-teleportation example, a linear dependence of $g_\textrm{opt}$ on $q$, more specifically $g_\textrm{opt} \approx 0.73 q + 0.27$ which represent a significant deviation from both standard choices $g=q$ and $g=1$. It should be kept in mind though that, like for the teleportation of the dual-rail qubit, $g_\textrm{opt}$ represents the optimal choice with respect to the fidelity without postselection. One may suspect that, when postselecting on the space of two dual-rail qubits (after the last beam splitter of the \textsc{csign} gate in Fig.~\ref{fig:KLMCZ}), the choice $g=q$ would yield a higher fidelity.

\begin{figure}
	\centering
	\includegraphics[height=0.52\columnwidth]{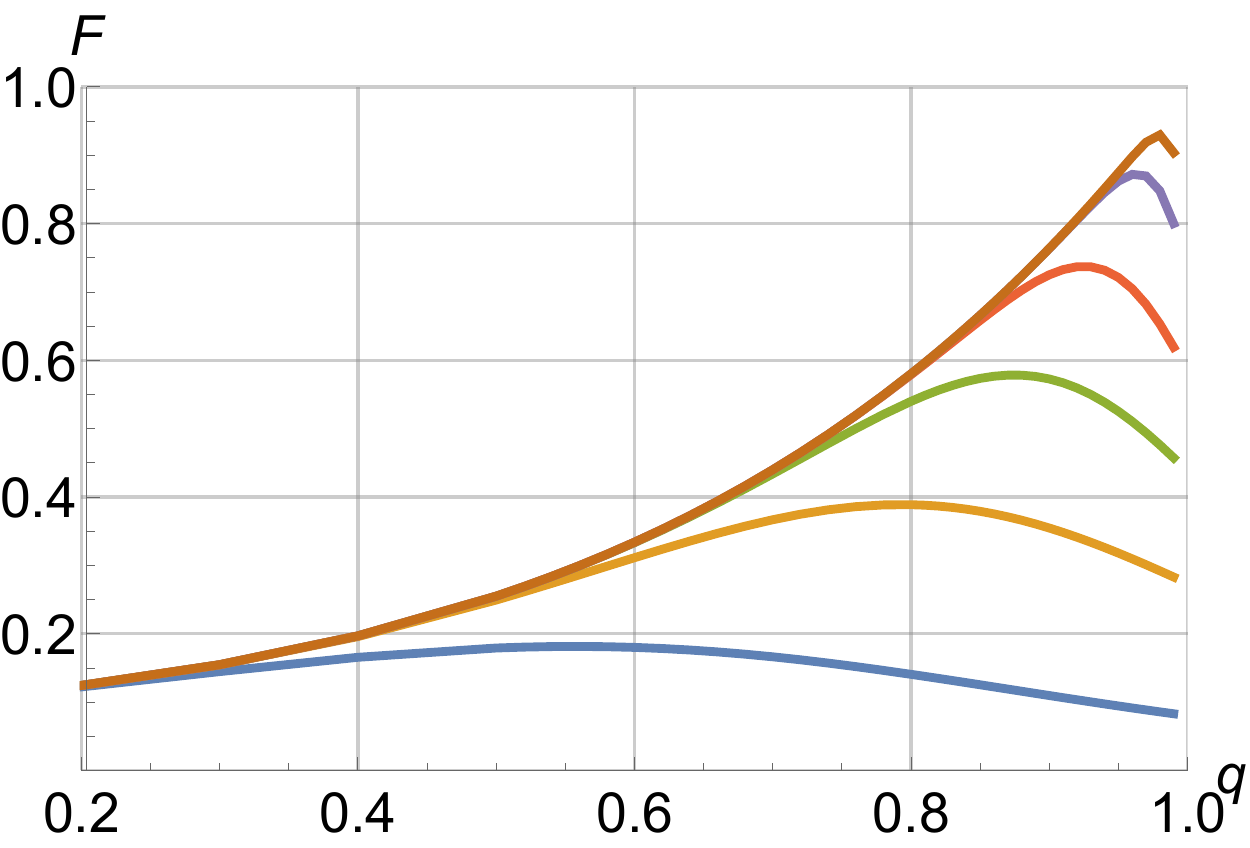}
	\caption{Worst-case fidelity of the \textsc{csign} gate as a function of the squeezing $q$. From bottom to top: $d=2,5,10,20,50,100$. Another version of this figure, i.e., Fig.~\ref{fig:CZ_fidelDB}, where the squeezing is represented in dB, can be found in Appendix~\protect\ref{sec:AppC}.}
	\label{fig:CZ_fidel}
\end{figure}

The worst-case fidelity obtained for the \textsc{csign} gate, using this optimized gain tuning, is shown in Fig.~\ref{fig:CZ_fidel}. There we clearly see that, when using a $\textsc{nss}_d$ gate with large enough $d$ in combination with an appropriate amount of squeezing, fidelities close to one are possible. The envelope of the functions actually indicates that, with increasing $d$, fidelities arbitrarily close to unit fidelity should be achievable. It should also be mentioned that the amount of squeezing required is within experimental reach. For example, the optimal amount of squeezing for the $d=100$ setup, which allows for a worst-case fidelity of $\sim 93\%$, is approximately $20$~dB.

This significant increase of the fidelity compared to the \textsc{nss} gate  teleportation with $d=2$, of course, comes at a price. While the CV Bell measurement is not affected by the choice of $d$ and remains deterministic, the creation of the teleportation resource state [see Eq.~\eqref{eq:cvres}] becomes highly probabilistic, and hence multiple attempts in parallel are required to achieve near-unit teleportation success probability. The cost of creating a two-mode squeezed state cannot be reduced to single-photon-source uses in a straightforward manner, but, in comparison to the $d-1$ single-photon sources required for the $\textsc{nss}_d$ gate, its contribution to the cost is nearly negligible. For simplicity, we thus assume that a single attempt to creating the two-mode squeezed state and to applying the gate $\textsc{nss}_d$ to it requires $d$ single-photon sources. Accordingly, the number of single-photon sources required to obtain a success probability $p_\textrm{CV}$ of the CV-based \textsc{csign} gate is given by
\begin{align}
	n_{\textrm{CV},d} = 2 d \frac{\log(1-\sqrt{p_\textrm{CV}})}{\log(1-p_d)},
\end{align}

\begin{table}[b]
\caption{Number of single-photon sources $n_{\textrm{CV},d}$ required to achieve the \textsc{csign} gate success probability $p_\textrm{CV}$ with the gates $\textsc{nss}_d$. For the latter, a success probability of $p_d=1/d^2$ is assumed. Additionally, in the second column, the maixmally achievable wort-case fidelity (see Fig.~\protect\ref{fig:CZ_fidel}) is given. \label{tab:n_CV}}
\begin{ruledtabular}
\begin{tabular}{r|r|rrrrrr}
$d$ & $F_d^{\max}$  & $p_\textrm{CV}=0.1$ & 0.5 & 0.75 & 0.9 & 0.99 & 0.999\\\hline 
2 & 0.18 & 6 & 18 & 28 & 42 & 74 & 106 \rule{0pt}{2.6ex}\\
5 & 0.39 & 94 & 301 & 493 & 728 & 1.3 & 1.9k\\
10 & 0.58 & 757 & 2.4k & 4.0k & 5.9k & 10.5k & 15.1k\\
20 & 0.74 & 6.0k & 19.6k & 32.1k & 47.5k & 84.6k & 121k\\
50 & 0.87 & 95.0k & 307k & 502k & 742k & 1.3M & 1.9M\\
100 & 0.93 & 760k & 2.5M & 4.0M & 5.9M & 10.6M & 15.2M
\end{tabular}
\end{ruledtabular}
\end{table}

\noindent when the $\textsc{nss}_d$ gate is used. The factor two and the square root are due to the fact that both $\textsc{nss}$ gate teleportations must succeed. Table~\ref{tab:n_CV} shows the required number of photons for the values of $d$ that were also used in Fig.~\ref{fig:CZ_fidel} and various \textsc{csign} success probabilities $p_\textrm{CV}$. There, we assume that an optical implementation of the $\textsc{nss}_d$ gates with a success probability of $1/d^2$ is available.

In summary, Secs.~\ref{sec:CVQT} and \ref{sec:CVGT} showed that a CV gate teleportation of the \textsc{nss} gate is, in principle, possible with fidelities close to unity, when a modified version of the \textsc{nss} gate, i.e., $\textsc{nss}_d$, is applied offline to a two-mode squeezed vacuum state (of experimentally achievable squeezing) and when an $f$-deformed displacement, i.e., the correction operation associated to the self-Kerr gate, is applied to the output states. Two of these teleported \textsc{nss} gates can be utilized to construct a \textsc{csign} gate with both fidelity and success probability above $90\%$ and an overhead of a few million single-photon sources. While this is less resource efficient than the DV approach presented in Sec.~\ref{sec:DVGT}, it still offers a significant improvement over the standard KLM scheme.

\section{Comparison: DV vs. CV}\label{sec:comparison}

In this section, we shall summarize the DV and CV teleportation-assisted optical \textsc{csign} gates described in Secs.~\ref{sec:DVGT} and \ref{sec:CVGT}, and compare them with respect to success probability, fidelity, and resource efficiency.

For the DV approach the main drawback is the limited success probability of both the linear optical \textsc{csign} gate and the DV teleportation, namely the limitation of the standard Bell measurement to a success probability of $1/2$. On the other hand, DV techniques offer, in principle, unit-fidelity operations. In the case of CV, the teleportation succeeds deterministically, provided that the required offline resource state is available. Here, the major problem is the deformation induced by the teleportation due to finite squeezing, which leads to poor fidelities. This is even amplified for gate teleportation, as the $\textsc{nss}$ gate transforms the two-mode squeezed teleportation resource rather drastically (it effectively truncates the state). In order to compare the two approaches, the quality $Q=p\cdot F$, which was introduced in Sec.~\ref{sec:CVQT}, seems a useful figure of merit.

We have shown in the previous sections that, for both DV and CV, methods exist that allow an increase of the \textsc{csign} gate's quality above $90\%$ and, in principle, even arbitrarily close to unity. For comparison, the quality of the standard $\textsc{csign}$ gate of KLM \cite{KLM} is $1/16$ (or $1/4$ with standard gate teleportation techniques). In the DV approach, as presented in Sec.~\ref{sec:DVGT}, a significant increase is made possible by quantum gate teleportation with advanced linear-optics Bell measurement schemes. Most notably, the combination of the Bell measurements presented in Refs.~\cite{Grice} and \cite{Ewert} enables a significant increase of the success probability while keeping the experimental costs small (especially, when compared to the advanced schemes of Ref.~\cite{KLM}). In the CV case, see Secs.~\ref{sec:CVQT} and \ref{sec:CVGT}, the fidelity of an \textsc{nss} gate teleportation is increased substantially by using an $f$-deformed displacement as correction operation and replacing the standard \textsc{nss} gate with a generalized variant, which more closely resembles a strong self-Kerr effect.

Either route towards a near-perfect linear-optics \textsc{csign} gate, i.e., fidelities as well as success probabilities close to unity, are associated with a considerable overhead of resource requirements. To achieve qualities above $90\%$, the DV \textsc{csign} gate requires a few thousand single-photon sources, and for the CV variant the cost is in the range of millions of single-photon sources. However, especially in the DV case, a small number of additional photons set in the right way can increase the quality substantially. For example, adding 8 single photons to the standard \textsc{csign} gate teleportation of KLM and using them for two $75\%$ success probability Bell measurements as in Ref.~\cite{Ewert} increases the quality from $0.25$ to $0.75^2=0.56$. Since, as shown in Fig.~\ref{fig:DV_res_creation}, thousands of single-photon sources are required to generate the linear cluster state $\ket{\textrm{res}}$  [see Eq.~\eqref{eq:DV_res_state}] reliably, the additional 8 photons hardly make any difference in the total cost.

In all of the above considerations, the cost of the photon-number-resolving detectors has been neglected. However, taking these into account is very straightforward, as basically in all schemes described in the previous sections, the number of photon-number-resolving detectors is the same as the number of single-photon sources. However, there is an important distinction to be made. While photon-number-resolving detectors are crucial for the advanced Bell measurement schemes, applied in the DV approach, the CV variant utilizes them only in the resource state generation phase, i.e., in the $\textsc{nss}_d$ gates. Should some new technique for implementing such a gate, or even a probabilistic self-Kerr gate, become available, or if the resource states were to be provided by entirely different means, the resource cost of the detectors in the CV teleportation-assisted \textsc{csign} gate would drop drastically, as it then only requires two, readily available, homodyne detectors.

\section{Conclusion}\label{sec:conclusion}

We have presented two routes to constructing high-fidelity, near-deterministic, linear optical \textsc{csign} gates with the help of quantum gate teleportation. One method applies discrete-variable techniques, in particular, recent results on near-deterministic Bell measurements, to circumvent the usual limitations of teleportation success probabilities. The other approach utilizes deterministic continuous-variable teleportation and a newly developed, generalized form of the nonlinear-sign-shift gate, in order to reduce the loss of fidelity due to finite squeezing and misaligned correction operations. In terms of resource consumption, the DV approach is currently much more efficient and beats the original scheme by KLM to a near-perfect linear optical \textsc{csign} gate by several orders of magnitude.

\appendix
\section{Detailed analysis of a continuous-variable state teleportation}
\label{sec:AppA}

Here we investigate the success probability and the fidelity of quantum state teleportation ($U=\mathbbm{1}$) with the CV teleportation scheme of Fig.~\ref{fig:CVGateTele} and a special focus on the option of conditioning the teleportation on the outcome of the CV Bell measurement. We assume that the input state to be teleported is a pure state with finite Fock expansion:
\begin{align}
	\ket{\psi_\textrm{in}} = \sum_{n=0}^N c_n \ket{n}.
\end{align}
The probability density of obtaining the result $\beta$ in the CV Bell measurement is given by
\begin{align}
	&p(\beta)\nonumber\\
	&= \operatorname{tr}_{1,2}\big[\ket{\beta_\textrm{hm}}\!\bra{\beta_\textrm{hm}} \operatorname{tr}_3\big(\ket{\psi_\textrm{in}}\!\bra{\psi_\textrm{in}}\!\otimes\!\ket{\textrm{TMSS}_q}\!\bra{\textrm{TMSS}_q}\big)\big]\nonumber\\
	&=\operatorname{tr}\left[T_q(\beta)\ket{\psi_\textrm{in}}\!\bra{\psi_\textrm{in}}T_q^\dagger(\beta)\right]\label{eq:pbeta}.
\end{align}
The matrix elements of the transfer operator $T_q(\beta)$ in the Fock basis shall be calculated next (see also \cite{Takeda13,Fuwa14}). We start with
\begin{align}
	&T_q(\beta)\ket{0} = \sqrt{\frac{1-q^2}{\pi}}D(q \beta) \sum_{n=0}^\infty q^n \ket{n} \braket{n|-\beta}\nonumber\\
	&=\sqrt{\frac{1-q^2}{\pi}}D(q \beta) \sum_{n=0}^\infty q^n e^{-\abs{\beta}^2/2} \frac{(-\beta)^n}{\sqrt{n!}}\ket{n} \nonumber\\
	&=\sqrt{\frac{1-q^2}{\pi}} e^{-(1-q^2)\abs{\beta}^2/2} D(q\beta) \ket{-q \beta}\nonumber\\
	&=\sqrt{\frac{1-q^2}{\pi}} e^{-(1-q^2)\abs{\beta}^2/2} \ket{0}.
\end{align}
Using this we obtain
\begin{align}
	&T_{k,l} := \braket{k|T_q(\beta)|l} \label{eq:Tkl}\\
	&= \sqrt{\frac{1-q^2}{\pi}} \bra{k} D(q \beta) \sum_{n=0}^\infty q^n \ket{n}\!\bra{n} D^\dagger(\beta) \frac{(a^\dagger)^l}{\sqrt{l!}}\ket{0}\nonumber\\
	&= \sqrt{\frac{1-q^2}{\pi l!}} \bra{k} D(q \beta) \sum_{n=0}^\infty q^n \ket{n}\!\bra{n} (a^\dagger+\beta^*)^l\ket{-\beta}\nonumber\\
	&= \sqrt{\frac{1-q^2}{\pi l!}} \bra{k} D(q \beta) \sum_{n=0}^\infty \sum_{m=0}^l \binom{l}{m} q^{n+m} (a^\dagger)^m \ket{n}\nonumber\\
	&\qquad\quad\bra{n} (\beta^*)^{l-m}\ket{-\beta}\nonumber\\
	&= \frac{1}{\sqrt{l!}}\bra{k} \sum_{m=0}^l \binom{l}{m} q^m (a^\dagger-q\beta^*)^m (\beta^*)^{l-m} T_q(\beta) \ket{0}\nonumber\\
	&= \frac{1}{\sqrt{l!}}\bra{k} [\beta^*+q(a^\dagger-q \beta^*)]^l T_q(\beta) \ket{0}\nonumber\\
	&= \sqrt{\frac{1-q^2}{\pi}} e^{-(1-q^2)\abs{\beta}^2/2} \binom{l}{k}\sqrt{\frac{k!}{l!}}q^k [(1-q^2)\beta^*]^{l-k}.\nonumber
\end{align}
Inserting the latter into Eq.~\eqref{eq:pbeta} and integrating over the circular area in the complex plane gives the probability of a measurement result with $\abs{\beta}\leq B$:
\begin{align}
	P(B) &= \int_{\abs{\beta}\leq B} p(\beta) d^2\beta \nonumber\\
	&= \sum_{n,m=0}^N c_n c^*_m \sum_{k=0}^{\min(n,m)} \int_{\abs{\beta}\leq B} T_{k,n} T_{k,m}^* d^2\beta. \label{eq:PB1}
\end{align}
Note that Eq.~\eqref{eq:Tkl} implies that the transfer operator $T_q(\beta)$ can only decrease the number of photons [$k\leq l$ in Eq.~\eqref{eq:Tkl}], hence the restriction on the sum over $k$ (which is extremely useful in numerical calculations). Furthermore, Eq.~\eqref{eq:Tkl} shows that $\operatorname{arg}(T_{k,l}) = (k-l) \operatorname{arg}(\beta)$, where $\operatorname\arg$ is the complex argument modulo $2\pi$. The choice of a circular area for the conditioning of the measurement results is thus quite advantageous, as the integrals in Eq.~\eqref{eq:PB1} will yield a non-zero result only if $\operatorname{arg}(T_{k,n} T_{k,m}^*) \propto (n-m) = 0$. In this case, we obtain
\begin{align}
	&\int_{\abs{\beta}\leq B} \abs{T_{k,n}}^2 d^2\beta\nonumber\\
	&=\binom{n}{k} \frac{q^{2k}(1-q^2)^{n-k}}{(n-k)!}\nonumber\\
	&\qquad \quad\frac{(1-q^2)^{n-k+1}}{\pi} \int_{\abs{\beta}\leq B} \abs{\beta}^{2(n-k)} e^{-(1-q^2)\abs{\beta}^2} d^2\beta\nonumber\\
	&=\binom{n}{k} \frac{q^{2k}(1-q^2)^{n-k}}{(n-k)!}\int_0^{(1-q^2)B^2} x^{n-k} e^{-x} dx. \label{eq:gamma1}
\end{align}
The integral over $x$ is the incomplete gamma function, which has a useful representation as
\begin{align}
	(n-k)!\left\{1-e^{-(1-q^2)B^2}\sum_{j=0}^{n-k} \frac{[(1-q^2)B^2]^j}{j!}\right\},
\end{align}
which then gives
\begin{align}
	P(B) &= 1-e^{-(1-q^2)B^2}\sum_{n=0}^N \abs{c_n}^2 \\
	&\qquad\sum_{k=0}^n \binom{n}{k} q^{2k} (1-q^2)^{n-k} \sum_{j=0}^{n-k} \frac{[(1-q^2)B^2]^j}{j!},\nonumber
\end{align}
and, using several identities of the binomial coefficient, we obtain the final result for the success probability
\begin{align}
	P(B) &= 1-e^{-(1-q^2)B^2}\sum_{n=0}^N \abs{c_n}^2  \nonumber\\
	&\qquad \sum_{j=0}^{n} \frac{[(1-q^2)B]^{2j}}{j!} \sum_{k=0}^{n-j} \binom{k+j-1}{k} q^{2k}.
\end{align}
Let us now turn to the calculation of the fidelity of the state teleportation. The state after teleportation with a CV Bell measurement result $\beta$ is given by
\begin{align}
	&\rho(\beta) = \frac{T_g(\beta) \ket{\psi_\textrm{in}}\!\bra{\psi_\textrm{in}} T_g^\dagger(\beta) }{\operatorname{tr}\left[T_g(\beta) \ket{\psi_\textrm{in}}\!\bra{\psi_\textrm{in}} T_g^\dagger(\beta) \right]}\nonumber\\
	&= \frac{1}{p(\beta)} D(\gamma) T_q(\beta) \ket{\psi_\textrm{in}}\!\bra{\psi_\textrm{in}} T_q^\dagger(\beta) D^\dagger(\gamma),
\end{align}
where we have used the abbreviation $\gamma = (g-q)\beta$. The fidelity for this single teleportation event, depending on $\beta$, is thus given by
\begin{align}
	&f(\beta) = \braket{\psi_\textrm{in}|\rho(\beta)|\psi_\textrm{in}}\nonumber\\
	&= \sum_{n,m,k,l=0}^N\!\frac{c_n c_m^* c_k^* c_l}{p(\beta)} \braket{k|D(\gamma)T_q(\beta)|n}\!\braket{m|T_q^\dagger(\beta) D^\dagger(\gamma)|l}\nonumber\\
	&= \sum_{n,m,k,l=0}^N\!\frac{c_n c_m^* c_k^* c_l}{p(\beta)} \sum_{r=0}^n \sum_{s=0}^m D_{k,r} T_{r,n} T_{s,m}^* D_{l,s}^*.
\end{align}
Here we introduced
\begingroup
\allowdisplaybreaks
\begin{align}
	&D_{k,r}:=\braket{k|D(\gamma)|r}\label{eq:dispmatelem}\\
	&= \frac{1}{\sqrt{k!}}\braket{0|a^k D(\gamma)|r} = \frac{1}{\sqrt{k!}}\braket{0|D(\gamma)(a+\gamma)^k|r}\nonumber\\
	&=\frac{1}{\sqrt{k!}}\sum_{t=0}^k \binom{k}{t}\braket{-\gamma|a^t|r}\gamma^{k-t}\nonumber\\
	&=\frac{1}{\sqrt{k!}}\sum_{t=0}^{\min(k,r)}\!\binom{k}{t} \braket{-\gamma|r-t} \sqrt{\frac{r!}{(r-t)!}}\gamma^{k-t}\nonumber\\
	&=\sqrt{\frac{r!}{k!}}\sum_{t=0}^{\min(k,r)}\!\binom{k}{t} \frac{\gamma^{k-t}(-\gamma^*)^{r-t}}{(r-t)!}e^{-\abs{\gamma}^2/2}\nonumber\\
	&=\sqrt{\frac{r!}{k!}} e^{-\abs{\gamma}^2/2} (-\gamma^*)^{r-k} \sum_{t=0}^{\min(k,r)}\!\binom{k}{t} \frac{\left(-\abs{\gamma}^2\right)^{k-t}}{(r-t)!}.\nonumber
\end{align}
\endgroup
Notably, the argument of $D_{k,r}$, just as that of $T_{k,l}$ [see Eq.~\eqref{eq:Tkl}], relates directly to the argument of $\beta$, as $\operatorname{arg}(D_{k,r}) = (k-r) \operatorname{arg}(\gamma) = (k-r) \operatorname{arg}(\beta)$. As the result of the homodyne measurement is random, we are interested in the fidelity averaged over all allowed values of $\beta$. We obtain
\begin{align}
	&F(B) = \int_{\abs{\beta}\leq B} f(\beta) p(\beta)  d^2\beta \label{eq:FB1}\\
	&= \sum_{n,m,k,l=0}^N\!c_n c^*_m c_k^* c_l \sum_{r=0}^n \sum_{s=0}^m \int\limits_{\abs{\beta}\leq B}\!D_{k,r} T_{r,n} T_{s,m}^* D_{l,s}^* d^2\beta. \nonumber 
\end{align}
By inserting the results from Eqs.~\eqref{eq:Tkl} and \eqref{eq:dispmatelem}, using $\operatorname{arg}(D_{k,r} T_{r,n} T_{s,m}^* D_{l,s}^*) = (k-n+m-l) \operatorname{arg}(\beta)$ to eliminate $l$ and after a lengthy calculation, this can be rewritten as
\begin{widetext}
\begin{align}
F(B) &= \sum_{n,m,k=0}^N\!c_n c^*_m c_k^* c_{k-n+m} \sqrt{\frac{n!m!}{k!(k-n+m)!}} \sum_{u=0}^{\substack{\min(k,n) + \\ \min(k-n+m,m)}} \frac{(1-q^2)q^u(g-q)^{2k-n+m-u}(1-gq)^{n+m-u}}{[(1-q^2)-(g-q)^2]^{m+k-u+1}}\nonumber\\
&\qquad \sum_{v=0}^{\min(k,n)} \binom{k}{v} \binom{k-n+m}{u-v} \frac{1}{(n-v)!(m-u+v)!} \int_0^{[(1-q^2)-(g-q)^2]B^2} x^{k+m-u} e^{-x} dx. \label{eq:fidelB}
\end{align}
\end{widetext}
The sum over $v$ is simply a combinatorical quantity and the integral is, once again, the incomplete gamma function.

\section{Generalization of the nonlinear-sign-shift gate}
\label{sec:AppB}

In this appendix we present a possible optical realization of the generalization of KLM's \textsc{nss} gate in order to more closely resemble the action of the strong self-Kerr effect $U_\textrm{SK}=\exp\left[i \tfrac{\pi}{2} \hat{n}(\hat{n}-1)\right]$ on Fock states. As described in the main text, the general setting is a (static) linear optics setup with a finite number of ancillary single photons and photon-number-resolving detectors. One should, for completeness, also add the restriction to a finite number of modes to the above properties. The interferometric part of the setup, which employs a unitary transformation of the creation operators of the involved modes, is characterized by the unitary matrix $u^{(d)}$ via
\begin{align}
	a_k^\dagger \rightarrow \sum_{l=1}^{d+E} u_{k,l}^{(d)} a_l^\dagger.
\end{align}
Here $d+E$ is the total number of modes used in our setup. In Ref.~\cite{Scheel04} it was shown that, in order to perform a non-trivial probabilistic gate in the fashion of KLM's \textsc{nss} gate acting on Fock states up to $d$, at least $d-1$ additional photons are required. Without loss of generality, we assume that these photons are injected in modes $2$ through $d$, while mode one is the input mode of the gate and the remaining $E$ modes are each prepared in the vacuum state. Since the generalized gate $\textsc{nss}_d$ is supposed to be diagonal in the Fock states, exactly $d-1$ photons must be directed to the detectors. Again, without loss of generality, we assume that these detections occur on the modes $2$ through $d$. For convenience, we further assume that a successful application of the gate is heralded by $d-1$ single-photon clicks in these $d-1$ detectors. Since all photons start and end up in the first $d$ modes in the case of success, the upper left $d\times d$-matrix of $u^{(d)}$, which we denote $v^{(d)}$, completely determines the action of the gate on the input mode. The remainder of the matrix $u^{(d)}$ is then solely needed to extend $v^{(d)}$ to a unitary matrix. This technique has also been employed in the development of the \textsc{csign} gate with a probability of $2/27$ in Ref.~\cite{Knill}, and it requires $\|v^{(d)}\|_2 \leq 1$.

In order to deduce an optical setup that realizes the generalized \textsc{nss} gate we start by defining the action of the gate on the Fock states in case of success as $\textsc{nss}_d \ket{n} = \alpha_{n}^{(d)} \ket{n}$. The $\alpha_{n}^{(d)}$ are functions of the matrix elements $v_{i,j}^{(d)}$. For example, in case of the standard \textsc{nss} gate we have for the vacuum state input $\alpha_0^{(2)} = v_{2,2}^{(2)}$, i.e., the single ancillary photon must end up in the detector of mode two. For the input state $\ket{1}$ we obtain $\alpha_1^{(2)} = v_{1,1}^{(2)} v_{2,2}^{(2)} + v_{1,2}^{(2)} v_{2,1}^{(2)}$. Of course, for larger setups and higher photon numbers, the terms become more involved. A useful point of view is that photons may scatter in and out of mode one. As the $\textsc{nss}_d$ gate is diagonal in the Fock states, the same number of photons must be scattered in as out, and we obtain
\begin{align}
	\alpha_n^{(d)} = \sum_{k=0}^{d-1} \binom{n}{k} \left[v_{1,1}^{(d)}\right]^{n-k} \beta_k^{(d)}, \label{eq:alphan}
\end{align}
where $\beta_k^{(d)}$ corresponds to the process that $k$ of the $d-1$ ancillary photons end up in mode one, while $k$ of the input photons end up in the detectors. Together with the conditions
\begin{align}
	\alpha_n^{(d)}=U_\textrm{SK}(n) \alpha_0^{(d)},\qquad \forall n\leq d,
\end{align}
where $U_\textrm{SK}(n) = \exp[i \frac{\pi}{2}n(n-1)]$, this yields a set of $d+1$ equations for the variables $\{\beta_0^{(d)},\ldots,\beta_{d-1}^{(d)},v_{1,1}^{(d)} \}$. One solution to this set of equations is given by
\begin{align}
	\beta_k^{(d)} &= \alpha_0^{(d)} \sum_{l=0}^{k} \binom{k}{l} \left[v_{1,1}^{(d)}\right]^l U_\textrm{SK}(k+l), \label{eq:betasol}\\
	v_{1,1}^{(d)} &= (-1)^{d+1} \tan\left(\tfrac{\pi}{4 d}\right).\label{eq:v11sol}
\end{align}
Multiple solutions for $v_{1,1}^{(d)}$ exist, but this is the most important one, as it is the smallest with respect to its absolute value and because it ensures that the gate suppresses higher Fock states exponentially and does not amplify them. As the choice of $\{\beta_0^{(d)},\ldots,\beta_{d-1}^{(d)},v_{1,1}^{(d)} \}$ completely determines the action of the gate on all Fock states via Eq.~\eqref{eq:alphan}, which is valid for all $n$, including $n>d$, it only remains to be shown that matrices $v^{(d)}$ exist with $\|v^{(d)}\|_2 \leq 1$ that give the correct values for $\beta_{k}^{(d)}$. For a proof of principle, we have devised a set of matrices
\begin{align}
	v^{(d)}= \begin{pmatrix}
		v_{1,1} & x & \cdots & \cdots & x\\
		\lambda_1^{(d)} & x & 0 & \cdots & 0\\
		\vdots & 0 & \ddots & \ddots & \vdots\\
		\lambda_{d-2}^{(d)} & \vdots & \ddots & x & 0\\
		\frac{\alpha_0^{(d)} \lambda_{d-1}^{(d)}}{x^{d-1}} & 0 & \cdots & 0 & \frac{\alpha_0^{(d)}}{x^{d-2}}
	\end{pmatrix}\label{eq:vmat}
\end{align}
that are tunable such that, by reducing the success probability $p_d = \abs{\alpha_0^{(d)}}^2$ sufficiently much, a value $x$ exists for which the condition $\|v^{(d)}\|_2 \leq 1$ is fulfilled. The variables $\{\lambda_1^{(d)},\ldots,\lambda_{d-1}^{(d)}\}$ are used to fulfill Eq.~\eqref{eq:betasol}. Due to the special form of the matrix, we can write
\begin{align}
	\beta_k^{(d)} = \alpha_0^{(d)} \sum_{\substack{\vec{l}\in \{1,...,d-1\}^k\\l_i\neq l_j \forall i\neq j}} \prod_{i=1}^k \lambda_{l_i}^{(d)}. \label{eq:betalambda}
\end{align}
The combination of Eqs.~\eqref{eq:betasol} and \eqref{eq:betalambda} gives a set of $d-1$ homogeneous polynomial equations in the $\lambda_i^{(d)}$. Bezout's theorem then implies that this set of equations has at most $(d-1)!$ solutions. However, since the equations are invariant under permutation of the $\lambda_i^{(d)}$, we know that if one solution exists, then there are $(d-1)!$ equivalent solutions, thus finding a single solution gives all solutions by permutation of the $\lambda_i^{(d)}$. It turns out that the most convenient solution is when the $\lambda_i^{(d)}$ are sorted in decreasing order with respect to their absolute value, in order to maximize the achievable success probability. Our numerical analysis has shown that, at least for setups with $d\leq 7$, solutions $\lambda_i^{(d)}$ exist that allow for a unitary extension of the matrix $v^{(d)}$. The resulting success probabilities are approximately given by $p_d \approx 25\cdot10^{-d}$. However, as mentioned in the main text, it seems very likely that, when not restricted to a special form of the matrix $v^{(d)}$ like in Eq.~\eqref{eq:vmat}, success probabilities with a much better scaling are achievable, potentially $p_d =1/d^2$.

\section{Detailed analysis of a continuous-variable gate teleportation}
\label{sec:AppC}

\begin{figure}[b]
	\centering
	\includegraphics[height=0.52\columnwidth]{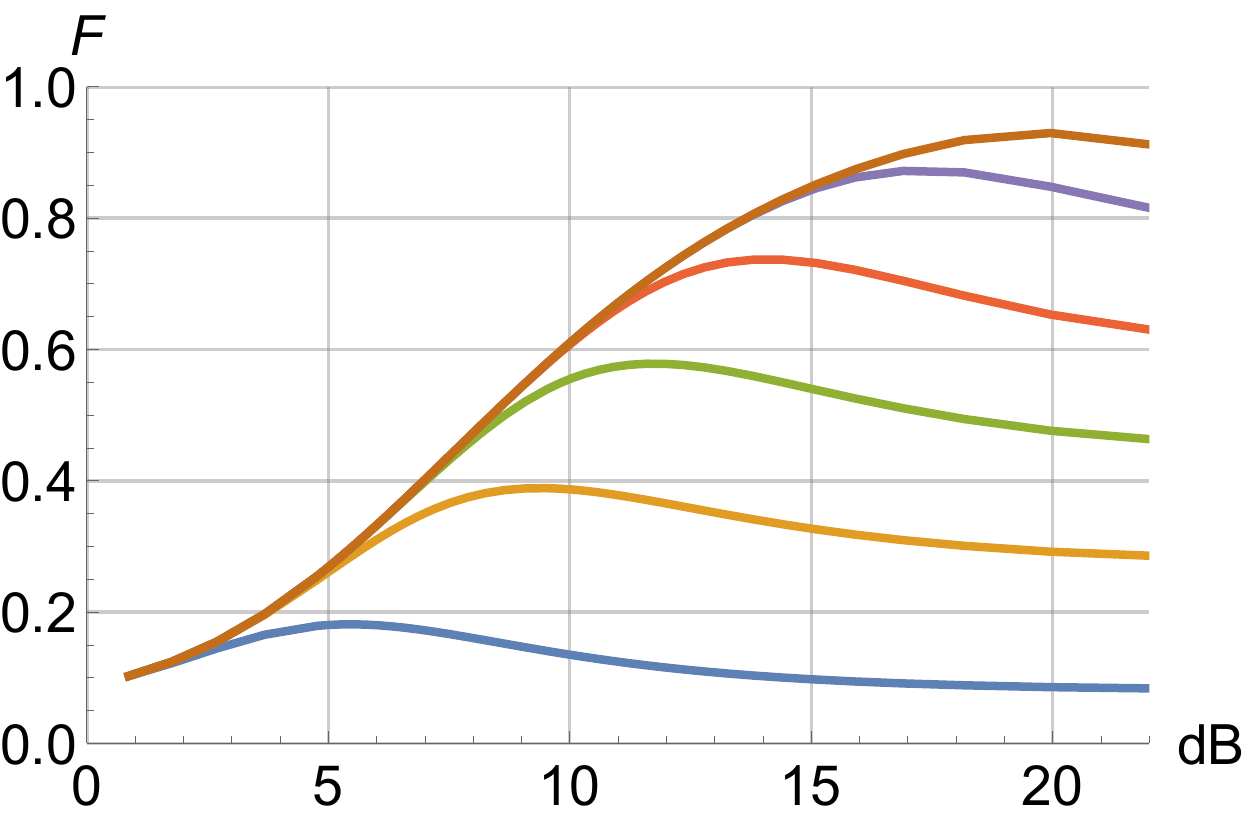}
	\caption{Worst-case fidelity of the \textsc{csign} gate as a function of the squeezing in dB. From bottom to top: $d=2,5,10,20,50,100$. This is another version of Fig.~\protect\ref{fig:CZ_fidel} in the main text, where the squeezing is represented by $q$.}
	\label{fig:CZ_fidelDB}
\end{figure}

Here we derive the formulae for the success probability and the fidelity of the \textsc{nss} gate teleportation with continuous variables. More general than in the main text, here we include the possibility of conditioning on the outcome of the CV Bell measurement. The following results have been used to verify that, for the gate teleportation just like for the state teleportation, the optimal quality $Q$ is obtained entirely without conditioning. Furthermore, we show how the gate teleportation fidelities of the two \textsc{nss} gates determine the fidelity of the \textsc{csign} gate. In addition, we give an alternative visual representation of the fidelity as a function of the squeezing: instead of the parameter $q$ its equivalent in dB is used in Fig.~\ref{fig:CZ_fidelDB}.

As mentioned in the main text, the central idea in order to obtain numerically tractable formulae is to truncate the two-mode squeezed resource state of the teleportation. This is consistent with the fact that the generalized gates $\textsc{nss}_d$ suppress higher Fock states exponentially. The resulting state after application of the $\textsc{nss}_d$ gate is thus given by
\begin{align}
	\mathcal{N}_{d,t}\ (\mathbbm{1}\otimes\textsc{nss}_d) \sum_{n=0}^{d+t} q^n \ket{n,n},
\end{align}
where $t$ represents the truncation limit and the normalizing coefficient $\mathcal{N}_{d,t}$ is given by
\begin{align}
	\frac{1}{\mathcal{N}_{d,t}^2} &= \sum_{n=0}^{d+t} q^{2n} \abs{\alpha_n^{(d)}}^2\nonumber\\
	&= \frac{1-q^{2(d+1)}}{1-q^2} + q^{2 d} \sum_{n=1}^t q^{2n}\abs{\alpha_{d+n}^{(d)}}^2,
\end{align}

\noindent where the values of $\alpha_n^{(d)}$ are given by Eqs.~\eqref{eq:alphan}, \eqref{eq:betasol}, and \eqref{eq:v11sol}. The derivation of the probability of obtaining a CV Bell measurement result with $\abs{\beta}\leq B$ is, in principle, the same as for the state teleportation case. However, the new transfer operator cannot be expressed in terms of $T_q(\beta)$. Hence, the matrix elements of displacement operators must be used. These are sufficient, since both the \textsc{nss} gate and the self-Kerr gate are diagonal in the Fock basis. We obtain
\begin{align}
	P_d(B) &= \mathcal{N}_{d,t}^2\sum_{n,m=0}^2 c_n c_m^* \sum_{k=0}^{d+t} q^{2k} \abs{\alpha_{k}^{(d)}}^2\nonumber\\
	&\qquad \int_{\abs{\beta}\leq B} D_{n,k}^*(\beta) D_{m,k}(\beta) d^2\beta,
\end{align}
where $D_{n,k}(\beta)$ is given by Eq.~\eqref{eq:dispmatelem} when replacing $\gamma$ with $\beta$ and where we have assumed the input state to contain no more than two photons. Again, the integration over the circular area yields $n=m$ and the same simplifications as in Appendix~\ref{sec:AppA} yield
\begin{align}
	P_d(B) &= \mathcal{N}_{d,t}^2 \sum_{n=0}^2\abs{c_n}^2\sum_{k=0}^{d+t} q^{2k} \abs{\alpha_{k}^{(d)}}^2\\
	&\qquad\sum_{l=0}^{2\min(n,k)} \gamma(k+n-l+1,B^2)\nonumber\\
	&\qquad\quad\sum_{j=0}^{\min(n,k)} \binom{n}{j}\binom{k}{l-j}\frac{(-1)^l}{(k-j)!(n-l+j)!},\nonumber
\end{align}
where the sum over $j$ is another combinatorical identity, and $\gamma(s,x)$ is the incomplete gamma function that we already encountered in Eqs.~\eqref{eq:gamma1} and \eqref{eq:fidelB}.

For the calculation of the fidelity, the same approach works leading to
\begin{align}
	&F_d(B)\nonumber\\
	&= \mathcal{N}_{d,t}^2 \sum_{n,m,k,l=0}^2\!c_n c^*_m c_k^* c_l \sum_{r,s=0}^{d+t} q^{r+s} U_\textrm{SK}(r)\alpha_r^{(d)}\alpha_s^{(d)}U_\textrm{SK}(s)\nonumber\\
	&\qquad\int_{\abs{\beta}\leq B}\!D_{k,r}(g\beta) D^*_{n,r}(\beta) D_{m,s}(\beta) D_{l,s}^*(g\beta) d^2\beta.
\end{align}
As it was the case for the state teleportation in Appendix~\ref{sec:AppA}, the integration over the circular area leads to $l=k-n+m$ and the usual simplifications give
\begin{widetext}
\begin{align}
	F_d(B) &= \mathcal{N}_{d,t}^2 \sum_{n,m,k=0}^2\!c_n c^*_m c_k^* c_l \sum_{r,s=0}^{d+t} q^{r+s} U_\textrm{SK}(r)\alpha_r^{(d)}\alpha_s^{(d)}U_\textrm{SK}(s) \frac{r! s!}{\sqrt{n!m!k!(k-n+m)!}}\nonumber\\
	&\qquad \sum_{u=0}^{\substack{\min(n,r) + \\ \min(m,s)}}\sum_{v=0}^{\substack{\min(k,r) + \\ \min(k-n+m,s)}} \frac{g^{r+s+n+m-2u}}{(1+g^2)^{r+s+m+k-u-v+1}} \gamma[r+s+m+k-u-v+1,(1+g^2)B^2]\nonumber\\
	&\qquad\sum_{w=0}^{\min(n,r)} \binom{n}{w}\binom{m}{u-w}\frac{(-1)^u}{(r-w)!(s-u+w)!} \sum_{z=0}^{\min(k,r)} \binom{k}{z}\binom{k-n+m}{v-z}\frac{(-1)^v}{(r-z)!(s-v+z)!}.\label{eq:fidelBd}
\end{align}
\end{widetext}
Due to the non-standard choice of the gain-tuning, $g\neq q$, and the misalignment of the $\textsc{nss}_d$ gate and the correction corresponding to the self-Kerr gate, further simplifications, which reduce the complexity of the expression for the fidelity substantially, are not available here. For example, if a unitary gate $U$ is to be teleported, which is diagonal in the Fock basis like $U_\textrm{SK}$, and its proper correction $C(g\beta) = U D(g \beta)U^\dagger$ is applied, the gate teleportation fidelity exactly coincides with the state teleportation fidelity. If, furthermore, the gain-tuning is set to the standard choice $g=q$, one finds that the gate teleportation is equivalent to an amplitude damping channel followed by the (perfectly executed) gate $U$.

Next, we show how the above results can be used to calculate the fidelity of the CV \textsc{csign} gate. To this end, we define
\begin{align}
	F^{(d)}_{n,m,k,l} \!=\! \int\limits_{\abs{\beta}\leq B}\!\braket{k|U_\textrm{SK}^\dagger\tilde{T}^{(d)}_g(\beta)|n}\!\braket{m|[\tilde{T}^{(d)}_g]^\dagger(\beta) U_\textrm{SK}|l} d^2 \beta,
\end{align}
where $\tilde{T}^{(d)}_g(\beta)$ is the transfer operator associated with the \textsc{nss} gate teleportation with the $\textsc{nss}_d$ gate applied to the resource state. Even though $\tilde{T}^{(d)}_g(\beta)$ has not been defined explicitly, it is clear that $F^{(d)}_{n,m,k,k-n+m}$ is one summand in Eq.~\eqref{eq:fidelBd}, i.e., $n$, $m$, and $k$ are fixed and the coefficients $c_i$ are ignored. The fidelity of the \textsc{csign} gate can easily be expressed in terms of these $F^{(d)}_{n,m,k,k-n+m}$. Here, we show this for the example of the input state $\ketb{11}$, since then the minimal fidelity and thus the worst-case fidelity of the \textsc{csign} gate is obtained. Furthermore, while the extension to include an arbitrary input state is straightforward, it is rather lengthy.

The total four-mode input state $\ketb{11}$ leads to an input state to the two \textsc{nss} gates of the form $\frac{1}{\sqrt{2}}(\ket{0,2}-\ket{2,0})$. Thus, the fidelity of the \textsc{csign} gate acting on the state $\ketb{11}$ can be written as
\begin{align*}
	&F_\textsc{csign}(\ketb{11}) = \label{eq:fcsign}\numberthis\\
	&\sum_{\substack{n,m,k,l \\\in\{0,2\} \\ n-k = m-l}}\frac{1}{4}\int\limits_{\abs{\beta_1}\leq B}\int\limits_{\abs{\beta_2}\leq B}\!\braket{k|\tilde{T}^{(d)}_{g}(\beta_1)|n}\!\braket{m|[\tilde{T}^{(d)}_{g}]^\dagger(\beta_1)|l} \\
	&\braket{2-k|\tilde{T}^{(d)}_{g}(\beta_2)|2-n}\!\braket{2-m|[\tilde{T}^{(d)}_{g}]^\dagger(\beta_2)|2-l} d^2\beta_2 d^2\beta_1\displaybreak\\
	&= \frac{1}{2}\left[F^{(d)}_{0,0,0,0} F^{(d)}_{2,2,2,2} + F^{(d)}_{0,0,2,2} F^{(d)}_{2,2,0,0} + F^{(d)}_{0,2,0,2} F^{(d)}_{2,0,2,0}\right].
\end{align*}

It may not be obvious from Eq.~\eqref{eq:fcsign}, but our numerical analysis showed that this complicated formula for the worst-case fidelity of the \textsc{csign} gate can be approximated quite well by the product of the fidelities for two separate \textsc{nss} gate teleportations of the Fock states $\ket{0}$ and $\ket{2}$, respectively.\vfill

\bibliography{optCSIGN}

\end{document}